\documentclass[aps]{revtex4}
\usepackage{amsmath}
\usepackage{color}
\usepackage{psfrag}
\usepackage{dcolumn}
\usepackage{bm}
\usepackage[latin1]{inputenc}
\usepackage[spanish,english]{babel}
\usepackage{amsfonts}
\usepackage{amssymb,epsf}
\usepackage{graphicx}
\usepackage{epstopdf}

\begin{document}

\title{\textbf{Charged Accelerating BTZ Black Holes}}
\author{B. Eslam Panah$^{1,2,3}$\footnote{
email address: eslampanah@umz.ac.ir}}
\affiliation{$^{1}$ Department of Theoretical Physics, Faculty of Science, University of
Mazandaran, P. O. Box 47415-416, Babolsar, Iran\\
$^{2}$ ICRANet-Mazandaran, University of Mazandaran, P. O. Box 47415-416,
Babolsar, Iran\\
$^{3}$ ICRANet, Piazza della Repubblica 10, I-65122 Pescara, Italy}

\begin{abstract}
In this paper, we first extract general uncharged accelerating BTZ black hole solutions and study some of their properties. Our analysis shows that spacetime's asymptotic behavior depends on four parameters: the cosmological constant, mass, acceleration, and topological constant. Then, we study the temperature of these black holes and find that the temperature is always positive for AdS spacetime. Next, we extend our study for extracting charged accelerating BTZ black hole solutions in the presence of a nonlinear electrodynamics field known as conformally invariant Maxwell. Our findings indicate a coupling between the electrical charge and other quantities of the accelerating BTZ black holes. The asymptotic behavior of charged accelerating BTZ black holes depends on five parameters: the cosmological constant, the electrical charge, mass, the acceleration parameter, and the topological constant. Then, we studied the effects of charge, acceleration parameters, and the topological constant on the root of these black holes. Finally, we investigate the temperature of these black holes in AdS spacetime. For these black holes, the temperature depends on the electrical charge, accelerating parameter, and the cosmological constant. Our analysis indicates that the temperature of charged accelerating BTZ AdS black holes is always positive.
\end{abstract}

\maketitle



\section{\noindent Introduction}

The first black hole solution in the $(2+1)$-dimensions with the negative cosmological constant has been obtained by Banados-Teitelboim-Zanelli \cite{BTZ1,BTZ2}, which is known as BTZ black hole. BTZ black hole helps us find a profound insight into the physics of black hole, the quantum view of gravity, and its relations to string theory \cite{Import1,Import2,Import3}. This black hole being locally AdS$_{3}$ without curvature singularity may be considered as a prototype for general AdS/CFT correspondence \cite{AdS1,AdS2}. So, the $(2+1)-$dimensional gravity has no local dynamics, but the black hole horizon induces an effective boundary. Therefore, quantum studies around BTZ black holes can help better understand AdS/CFT correspondence. Another motivation for studying BTZ black holes is related to the salient problems of quantum black holes, such as loss of information and the endpoint of quantum evaporation, which can be more easily understood in some simple low-dimensional models than directly in four dimensions \cite{Harvey}. In addition, BTZ black holes are related via T-duality and U-duality to classes of asymptotically flat black strings \cite{Horowitz1,Horowitz2}.

Further, other attractive black hole solutions are related to accelerating black holes (which could evolve into supermassive black holes \cite{massiveI,massiveII}). They have been extracted in the $C-$metric \cite{Kinnersley,Plebanski,Dias,Griffiths}. These black holes have a conical deficit angle along one polar axis, providing the force driving the acceleration. It is noteworthy that the conical singularity pulling the black hole can be replaced by a magnetic flux tube \cite{Dowker1994} or a cosmic string \cite{Gregory1995}. One can imagine that something similar to the $C-$metric with its distorted horizon could explain a black hole that has been accelerated by an interaction with a local cosmological medium. It was indicated that the Hawking temperature of accelerating black holes is more than the Unruh temperature of the accelerated frame \cite{Letelier}. Thermodynamics and phase transitions of accelerating black holes have been evaluated in Refs. \cite{Appels,AppelsII,LiuM,Marco,Anabalon,Abbasvandi,AnabalonII,Ball,Jafarzade1}. Considering the charged accelerating black holes, the late-time growth rate of complexity differs from the ordinary charged black holes \cite{Jiang}. The efficiency of heat engine for (an)a (un)charged accelerating non-rotating AdS black hole \cite{Zhang2018,Zhang2019}, charged accelerating rotating black hole \cite{Jafarzade,Ahmed}, have been studied. The results showed that these black holes' parameters could influence heat engine efficiency (see Refs. \cite{GregoryI,GregoryII}, for more details). Quasinormal modes and late-time tails of the accelerating black holes have been studied in Ref. \cite{Destounis}. The causal structure of accelerating black holes has been discussed in Refs. \cite{CausalI,CausalII}. Geodesics and gravitational lensing in the spacetime of these black holes have been studied in Refs. \cite{GeodesicI,GeodesicII}. The effect of acceleration on the orbit of photons and the shadow of black holes have been evaluated in Refs. \cite{Grenzebach,ZhangJ}. It was indicated that the circular orbits of the photons deviated from the equatorial plane. Also, the property of the shadow of black holes changed due to the existence of acceleration. The $(2+1)-$dimensional uncharged accelerating black holes have been obtained in Refs. \cite{Astorino,Xu1,Xu2}. Then, the accelerating systems in $(2+1)-$dimensional spacetime have been evaluated in Ref. \cite{Henriquez}, and three exciting classes of geometry by studying holographically their physical parameters have been found.

The coupling of GR and nonlinear sources attracted significant attention because of their specific and exciting properties. One of these nonlinear theories in physics is nonlinear electrodynamics (NED). NED comes from the fact that these theories generalize the Maxwell field. Other motivations for considering NED are related to limitations of the Maxwell theory \cite{Delphenich1}, the radiation propagation inside specific materials \cite{Lorenci,Novello1}, and also, the description of the self-interaction of virtual electron-positron pairs \cite{Schwinger,Yajima}. In addition, it was shown that NED objects could remove both black holes and big bang singularities \cite{Ayon-Beato,Lorenci2,Dymnikova,Corda}. On the other hand, the electric field of a point-like charge encounters a singularity in its location (origin). In order to remove this divergence, Born and Infeld have introduced a NED, known as Born--Infeld (BI) theory \cite{Born}. One of the special classes of NED sources is related to the power Maxwell invariant (PMI). The Lagrangian of PMI is an arbitrary power of Maxwell Lagrangian \cite{PMI1,PMI2,PMI3,PMI5,PMI6,PMI7}. This model is considerably richer than Maxwell's theory because; i) it reduces to a linear Maxwell field in the special case (unit power). ii) PMI NED can remove the singularities of the electric fields of point-like charges at their locations \cite{singuPMI}. iii) it is a conformal invariance when the power of the Maxwell invariant is a quarter of spacetime dimensions (power=dimensions/4). In this case, the energy-momentum tensor is traceless, leading to conformal invariance. This idea is related to taking advantage of the conformal symmetry to construct the analogs of the $(3+1)-$dimensional Reissner-Nordstrom black holes with an inverse square electric field in arbitrary dimensions \cite{PMI1,Conf2}.

According to the mentioned important reasons for $(2+1)-$dimensional and
accelerating black holes and the exciting properties of NED, we are
interested in finding exact charged accelerating BTZ black hole solutions in
this paper. In other words, by coupling GR and NED, we intend to extract
exact charged accelerating BTZ black hole solutions in the presence of the
cosmological constant.

The structure of this paper is as follows: at first, we obtain the $(2+1)-$
dimensional uncharged accelerating black holes. Then, we show that the
obtained solutions recover the known BTZ black holes. In addition, we study
some properties of these solutions. Next, we extract the charged
accelerating BTZ black holes in the presence of the PMI field. We indicate
that these solutions turn to the known BTZ black holes in
Einstein-conformally invariant Maxwell gravity. Then, we study some exciting
properties of these solutions. The final section is devoted to concluding
remarks. 

\section{\noindent Uncharged Accelerating BTZ Black Hole Solutions}

This section devotes to the accelerating BTZ black holes without considering the matter field. The $(2+1)-$dimensional action in the Einstein-$\Lambda $ gravity can be written as 
\begin{equation}
\mathcal{I}(g_{\mu \nu },A_{\mu })=\frac{1}{16\pi }\int_{\partial \mathcal{M}%
}d^{3}x\sqrt{-g}\left[ R-2\Lambda \right] ,  \label{Action1}
\end{equation}%
where $R$ and $\Lambda $ are the Ricci scalar and the cosmological constant, respectively.

Varying the action (\ref{Action1}), with respect to the gravitational field $g_{\mu \nu }$, yields 
\begin{equation}
G_{\mu \nu }+\Lambda g_{\mu \nu }=0.  \label{Eq1}
\end{equation}

To obtain the accelerating BTZ black hole, we consider a $(2+1)-$dimensional spacetime introduced in Refs. \cite{Xu1,Xu2} 
\begin{equation}
ds^{2}=\frac{1}{\mathcal{K}^{2}\left( r,\theta \right) }\left[ -f(r,\theta
)dt^{2}+\frac{dr^{2}}{f(r,\theta )}+r^{2}d\theta ^{2}\right] ,
\label{Metric}
\end{equation}%
where $\mathcal{K}\left( r,\theta \right) =\alpha r\cosh \left( \sqrt{m-k}%
\theta \right) -1$, which is called the conformal factor. Also, $\alpha $ is
related to the acceleration parameter, and $k$ is a topological constant
that can be $\pm 1$ or $0$ \cite{Xu1,Topological1,Topological2}. The
coordinates of $t$ and $r$, respectively, are in the ranges $-\infty
<t<\infty $ and $0\leq r<\infty $. Due to the lack of translational symmetry
in $\theta $ (in the presence of the conformal factor), we seem to have no
reason to restrict it to be in $\left[ -\pi ,\pi \right] $. Therefore we
consider $\theta $ in the range $-\pi \leq \theta \leq \pi $. We will see
later that in some cases the range for $\theta $ still needs to be adjusted
for observers living on one side of the conformal infinity. The necessity of
adjusting the range of $\theta $ in certain cases is one of the major points
we will discuss in the following section. Also, $f(r,\theta )$ is a metric
function, which we have to find it.

Now, we should find a suitable metric function $f(r,\theta )$ to satisfy all components of Eq. (\ref{Eq1}). Considering the metric (\ref{Metric}), one can show that Eq. (\ref{Eq1}) turns to 
\begin{eqnarray}
Eq_{tt} &=&Eq_{rr}=2\alpha ^{2}r^{2}\left( ff^{\mathcal{\theta \theta }}-%
\frac{3}{2}f^{\mathcal{\theta }^{2}}\right) \cosh ^{2}\left( \sqrt{m-k}%
\theta \right)  \notag \\
&&  \notag \\
&&-2\alpha r\cosh \left( \sqrt{m-k}\theta \right) \left[ \alpha rf\sqrt{m-k}%
\sinh \left( \sqrt{m-k}\theta \right) f^{\mathcal{\theta }}+2ff^{\mathcal{%
\theta \theta }}-3f^{\mathcal{\theta }^{2}}-2f^{2}\left( \frac{rf^{\prime }}{%
2}-f+k-m\right) \right]  \notag \\
&&  \notag \\
&&+2\alpha rf\sqrt{m-k}\sinh \left( \sqrt{m-k}\theta \right) f^{\mathcal{%
\theta }}+2ff^{\mathcal{\theta \theta }}-3f^{\mathcal{\theta }^{2}}-4rf^{2}%
\left[ \frac{f^{\prime }}{2}+\left( \Lambda +\left( k-m\right) \alpha
^{2}\right) r\right] ,  \label{Eq1a} \\
&&  \notag \\
Eq_{\theta \theta } &=&\alpha ^{2}r^{2}\cosh ^{2}\left( \sqrt{m-k}\theta
\right) \left[ 2r^{2}f^{2}f^{\prime \prime }-f^{\mathcal{\theta }%
2}-4f^{2}\left( k-m+rf^{\prime }-f\right) \right] +2r^{2}f^{2}f^{\prime
\prime }-f^{\mathcal{\theta }^{2}}  \notag \\
&&  \notag \\
&&+\alpha r\cosh \left( \sqrt{m-k}\theta \right) \left[ 4rf^{2}f^{\prime
}+2f^{\mathcal{\theta }^{2}}-4r^{2}f^{2}f^{\prime \prime }\right]
+4r^{2}f^{2}\left( \Lambda +\left( k-m\right) \alpha ^{2}\right) ,
\label{Eq1b} \\
&&  \notag \\
Eq_{\theta r} &=&Eq_{r\theta }=\alpha r\cosh \left( \sqrt{m-k}\theta \right) %
\left[ 2f^{\mathcal{\theta }}-rf^{\prime \mathcal{\theta }}\right]
+rf^{\prime \mathcal{\theta }}-f^{\mathcal{\theta }},  \label{Eq1c}
\end{eqnarray}%
where $f=f\left( r,\theta \right) $, $f^{\prime }=\frac{\partial f(r,\theta )%
}{\partial r}$, $f^{\prime \prime }=\frac{\partial ^{2}f(r,\theta )}{%
\partial r^{2}}$, $f^{\mathcal{\theta }}=\frac{\partial f(r,\theta )}{%
\partial \theta }$, $f^{\mathcal{\theta \theta }}=\frac{\partial
^{2}f(r,\theta )}{\partial \theta ^{2}}$, and $f^{\prime \mathcal{\theta }}=%
\frac{\partial ^{2}f(r,\theta )}{\partial \theta \partial r}$. It is notable
that $Eq_{tt}$, $Eq_{rr}$, $Eq_{\theta \theta }$ and $Eq_{\theta r}$ ($%
Eq_{r\theta }$)\ are related to components of $tt$, $rr$, $\theta \theta $\
and $\theta r$ ($r\theta $)\ of the Eq. (\ref{Eq1}). Finding an exact
solution from the above equations is very difficult. For this purpose we
consider $\theta =0$ in the equations (\ref{Eq1a})-(\ref{Eq1c}). This choice
reduces them to the following forms 
\begin{eqnarray}
Eq_{tt} &=&Eq_{rr}=2\alpha ^{2}\left( k-m\right) r+2\alpha \left[ f\left(
r\right) +m-k-\frac{rf^{\prime }\left( r\right) }{2}\right] +2r\Lambda
+f^{\prime }\left( r\right) ,  \label{Eq11a} \\
&&  \notag \\
Eq_{\theta \theta } &=&\left( \alpha r-1\right) \left[ \left( \alpha
r-1\right) f^{\prime \prime }\left( r\right) -2\alpha f^{\prime }\left(
r\right) \right] +2\left( \alpha ^{2}f\left( r\right) +\Lambda \right) ,
\label{Eq12a}
\end{eqnarray}%
where $Eq_{\theta r}=Eq_{r\theta }=0$, and also $f^{\prime }\left( r\right) =%
\frac{df\left( r\right) }{dr}$, and $f^{\prime \prime }\left( r\right) =%
\frac{d^{2}f\left( r\right) }{d^{2}r}$.

After some calculations, we find an exact general solution of Eqs. (\ref%
{Eq11a}) and (\ref{Eq12a}), in the following form 
\begin{equation}
f\left( r\right) =\left( \alpha r-1\right) ^{2}C\left( m,k,\Lambda ,\alpha
\right) +2\left( \alpha r-1\right) \left( m-k\right) -\left( \frac{2\alpha
r-1}{\alpha ^{2}}\right) \Lambda ,  \label{f(r)Uch}
\end{equation}%
where $C\left( m,k,\Lambda ,\alpha \right) $ is a constant which depends on
mass ($m$), the topological parameter ($k$), the cosmological constant ($%
\Lambda $), and the acceleration parameter ($\alpha $). We should note that
the solution (\ref{f(r)Uch}) satisfies all components of the field equation (%
\ref{Eq1}), provided we consider $\theta =0$.

In order to more investigate the obtained solution, we compare it with the
obtained solution in Ref. \cite{Xu1}. Considering $C\left( m,k,\Lambda
,\alpha \right) $ in the following form 
\begin{equation}
C\left( m,k,\Lambda ,\alpha \right) =m-k-\frac{\Lambda }{\alpha ^{2}},
\label{C1}
\end{equation}%
the solution (\ref{f(r)Uch}) reduces to 
\begin{equation}
f\left( r\right) =\left( \left( m-k\right) \alpha ^{2}-\Lambda \right)
r^{2}-m+k,  \label{f(r)C1}
\end{equation}%
in which covers the introduced solution in Refs. \cite{Xu1,Xu2}. Also, in the absence of the accelerating parameter ($\alpha =0$), the solution (\ref{f(r)C1}) turns to the well-known BTZ solutions as 
\begin{equation}
f\left( r\right) _{BTZ}=-\Lambda r^{2}-m+k,
\end{equation}

Let us recall the explanation of the conformal factor $1/K^{2}\left(
r,\theta \right) $\ in the metric (\ref{Metric}). Considering a static
observer in the following form 
\begin{equation}
x^{\mu }\left( \lambda \right) =\left( \lambda \frac{\mathcal{K}\left(
r,\theta \right) }{\sqrt{f\left( r\right) }},r,\theta \right) ,
\end{equation}%
in the spacetime, $\lambda $ is the proper time. Proper velocity is defined by $u^{\mu }=\frac{dx^{\mu }}{d\lambda }$, which is given by 
\begin{equation}
u^{\mu }\left( \lambda \right) =\left( \frac{\mathcal{K}\left( r,\theta
\right) }{\sqrt{f\left( r\right) }},0,0\right) ,
\end{equation}%

Straightforward calculations indicate that the proper acceleration $a^{\mu}=u^{\nu }\nabla _{\nu }u^{\mu }$ (for this kind of observer at $r=0$) leads to \cite{Xu1,Xu2} 
\begin{equation}
a^{\mu }a_{\mu }=\left( k-m\right) \alpha ^{2},  \label{alpha}
\end{equation}%
in which shows, $\alpha $ is proportional (but not equal) to the magnitude $\left\vert a\right\vert $ of the proper acceleration at the origin when $m\neq k$. Therefore, $\alpha $ may be called an acceleration parameter. It is worthwhile to mention that the $t-$component of $a^{\mu }$ is zero. So, $a^{\mu }$ is space-like, and we ought to have $a^{\mu }a_{\mu }>0$ in the static region of spacetime. Considering the equation (\ref{alpha}) for $m>k$, we see that $r=0$ is not in the static region. That the region of spacetime containing the origin $r=0$ is non-static is typical for black holes centered at the origin. For this reason, we infer that an accelerating black hole can exist only for $m>k$ (see Ref. \cite{Xu1}, for more details).

In the following, we study the properties of uncharged accelerating BTZ
black hole solutions (\ref{f(r)C1}) such as Ricci and Kretschmann scalars,
horizon structure, and temperature of these black holes. In other words, we
will evaluate the effect of various parameters of the accelerating BTZ black
holes on these quantities.

\subsection{Curvature Scalars}

Curvature invariant is scalar quantities constructed from tensors that
represent curvature. Two well-known curvature invariants are the Ricci and
Kretschmann scalars. The Ricci scalar curvature is defined as the trace of
the Ricci curvature tensor with respect to the metric. Also, the Kretschmann
scalar is constructed from the norm of the Riemann curvature tensor $%
R_{\alpha \beta \gamma \delta }$. To find the curvature singularity(ies) of
the spacetime, we calculate the Ricci and Kretschmann scalars. Using the
metric (\ref{Metric}) and after some algebraic manipulation, one can find
the Ricci scalar ($R$) and Kretschmann scalar ($R_{\alpha \beta \gamma
\delta }R^{\alpha \beta \gamma \delta }$) in the following forms 
\begin{eqnarray}
R &=&4\alpha ^{2}\left( \left( k-m-\frac{f\left( r\right) }{2}+\frac{%
rf^{\prime }\left( r\right) }{2}-\frac{r^{2}f^{\prime \prime }\left(
r\right) }{4}\right) \cosh ^{2}\left( \sqrt{m-k}\theta \right) +\frac{%
3\left( k-m\right) \sinh ^{2}\left( \sqrt{m-k}\theta \right) }{2}\right) 
\notag \\
&&  \notag \\
&&+\frac{2\alpha \cosh \left( \sqrt{m-k}\theta \right) }{r}\left(
k-m-f\left( r\right) +\frac{r^{2}f^{\prime \prime }\left( r\right) }{2}%
\right) -\frac{2f^{\prime }\left( r\right) +rf^{\prime \prime }\left(
r\right) }{r},  \label{RUnch1} \\
&&  \notag \\
R_{\alpha \beta \gamma \delta }R^{\alpha \beta \gamma \delta } &=&4\left[
r^{2}\mathcal{B}_{1}+\left( k-m\right) \mathcal{B}_{2}\sinh ^{2}\left( \sqrt{%
m-k}\theta \right) \right] \alpha ^{4}+16r\cosh ^{3}\left( \sqrt{m-k}\theta
\right) \left[ \tanh ^{2}\left( \sqrt{m-k}\theta \right) \mathcal{B}_{3}-%
\mathcal{B}_{4}\right] \alpha ^{3}  \notag \\
&&  \notag \\
&&+\left[ \frac{8\cosh ^{2}\left( \sqrt{m-k}\theta \right) \mathcal{B}_{5}}{%
r^{2}}-\frac{4\left( k-m\right) \sinh ^{2}\left( \sqrt{m-k}\theta \right)
\left( 2f^{\prime }\left( r\right) +rf^{\prime \prime }\left( r\right)
\right) }{r}\right] \alpha ^{2}  \notag \\
&&  \notag \\
&&-\frac{4\cosh \left( \sqrt{m-k}\theta \right) \alpha \mathcal{B}_{6}}{r^{2}%
}+f^{\prime \prime ^{2}}\left( r\right) +\frac{2f^{\prime ^{2}}\left(
r\right) }{r^{2}},  \label{KUnCh1}
\end{eqnarray}%
where $\mathcal{B}_{1}$, $\mathcal{B}_{2}$, $\mathcal{B}_{3}$, $\mathcal{B}%
_{4}$, $\mathcal{B}_{5}$\ and $\mathcal{B}_{6}$ are 
\begin{eqnarray}
\mathcal{B}_{1} &=&f^{2}(r)\left( \frac{1}{r^{2}}+\frac{f^{\prime \prime
}\left( r\right) -\frac{2f^{\prime }\left( r\right) }{r}}{f\left( r\right) }%
\right) +f^{\prime \prime ^{2}}\left( r\right) \left( \frac{r^{2}}{4}-\frac{%
rf^{\prime }\left( r\right) }{f^{\prime \prime }\left( r\right) }\right)
+f^{\prime ^{2}}\left( r\right) +\frac{2\left( k-m\right) ^{2}}{r^{2}}, 
\notag \\
&&  \notag \\
\mathcal{B}_{2} &=&\left[ 4\left( m-k\right) +2\left( rf^{\prime }\left(
r\right) -f\left( r\right) \right) -r^{2}f^{\prime \prime }\left( r\right) %
\right] \cosh ^{2}\left( \sqrt{m-k}\theta \right) +3\left( k-m\right) \sinh
^{2}\left( \sqrt{m-k}\theta \right) ,  \notag \\
&&  \notag \\
\mathcal{B}_{3} &=&\frac{\left( k-m\right) }{r^{2}}\left( k-m-f\left(
r\right) +\frac{r^{2}f^{\prime \prime }\left( r\right) }{2}\right) ,  \notag
\\
&& \\
\mathcal{B}_{4} &=&\frac{\left( f^{\prime \prime }\left( r\right) -\frac{%
f^{\prime }\left( r\right) }{r}\right) f\left( r\right) }{2}+\frac{\left(
k-m\right) \left( \frac{rf^{\prime }\left( r\right) }{2}-f\left( r\right)
+k-m\right) }{r^{2}}+\frac{r^{2}\left( 1-\frac{3f^{\prime }\left( r\right) }{%
rf^{\prime \prime }\left( r\right) }+\frac{2f^{\prime ^{2}}\left( r\right) }{%
r^{2}f^{\prime \prime ^{2}}\left( r\right) }\right) f^{\prime \prime
^{2}}\left( r\right) }{4},  \notag \\
&&  \notag \\
\mathcal{B}_{5} &=&\left( 1+\frac{r^{2}f^{\prime \prime }\left( r\right)
-2rf^{\prime }\left( r\right) }{2f\left( r\right) }\right) f^{2}(r)+\frac{%
3r^{2}\left( 1-\frac{2f^{\prime }\left( r\right) }{rf^{\prime \prime }\left(
r\right) }+\frac{f^{\prime ^{2}}\left( r\right) }{r^{2}f^{\prime \prime
^{2}}\left( r\right) }\right) f^{\prime \prime ^{2}}\left( r\right) }{4}%
+\left( k-m\right) \left[ k-m+2r\left( f^{\prime }\left( r\right) -\frac{%
f\left( r\right) }{r}\right) \right] ,  \notag \\
&&  \notag \\
\mathcal{B}_{6} &=&r^{3}f^{\prime \prime ^{2}}\left( r\right)
-r^{2}f^{\prime \prime }\left( r\right) f^{\prime }\left( r\right)
+rf^{\prime ^{2}}\left( r\right) -2f^{\prime }\left( r\right) \left[ f\left(
r\right) +m-k\right] ,  \notag
\end{eqnarray}%
where in the absence of acceleration parameter (i.e., $\alpha =0$) the Ricci and Kretschmann scalars reduce to 
\begin{eqnarray}
R &=&-\frac{2f^{\prime }\left( r\right) +rf^{\prime \prime }\left( r\right) 
}{r},  \notag \\
&& \\
R_{\alpha \beta \gamma \delta }R^{\alpha \beta \gamma \delta } &=&f^{\prime
\prime ^{2}}\left( r\right) +\frac{2f^{\prime ^{2}}\left( r\right) }{r^{2}}.
\notag
\end{eqnarray}

Considering the metric function (\ref{f(r)C1}) and the equations (\ref%
{RUnch1}) and (\ref{KUnCh1}), we obtain the Ricci and Kretschmann
scalars which are 
\begin{eqnarray}
R &=&6\Lambda ,  \label{RiccUch} \\
&&  \notag \\
R_{\alpha \beta \gamma \delta }R^{\alpha \beta \gamma \delta } &=&12\Lambda
^{2},  \label{KreshUch1}
\end{eqnarray}%
However, there is no singularity, but we encounter a conical singularity at $r=0$.

According to Eq. (\ref{f(r)C1}), the asymptotical behavior of the obtained
solution is dependent on the cosmological constant ($\Lambda $), the
parameter of acceleration ($\alpha $), mass ($m$) and topological parameter (%
$k$). As a result, the obtained spacetime is not exactly asymptotically AdS
due to the existence of different parameters in the asymptotical behavior of
spacetime.

\subsection{Horizon Structure}

Another quantity we are interested in studying is the black hole's horizon. In the previous section, we found a conical singularity at $r=0$. Here, our analysis indicates that there are two roots for the obtained metric function $f(r)$, which are in the following form 
\begin{equation}
r_{\pm }=\pm \sqrt{\frac{m-k}{\left( m-k\right) \alpha ^{2}-\Lambda }}.
\label{rootuch}
\end{equation}%
where $r_{+}$ and $r_{-}$ belong to positive and
negative roots, respectively.

Now we are in a position to evaluate the obtained real root. It is necessary to mention that we must respect to signatures of spacetime (\ref{Metric}) and also for recovering the static BTZ black hole, we have to restrict $m>k$, which leads to $\Lambda <\left( m-k\right) \alpha ^{2}$. In other words, our calculations indicate that we only encounter one real root for the uncharged accelerating BTZ black hole when $\Lambda <\left( m-k\right) \alpha ^{2}$.

To determine the asymptotical de Sitter (dS) or anti-de Sitter (AdS) of an accelerating BTZ black hole, we can re-express the cosmological constant by using Eq. (\ref{rootuch}) in terms of $r_{+}$ as

\begin{equation}
\Lambda =\left( m-k\right) \left( \frac{\alpha ^{2}r_{+}^{2}-1}{r_{+}^{2}}%
\right) ,  \label{Lambda}
\end{equation}%
where indicates that there are three cases.

i) $\alpha ^{2}r_{+}^{2}>1$, we encounter with dS accelerating black holes.

ii) the asymptotical behavior of accelerating BTZ black hole is AdS when $%
\alpha ^{2}r_{+}^{2}<1$.

iii) flat case is $\alpha ^{2}r_{+}^{2}=1$.

Recently it was shown that in three-dimensional dS spacetime, classical
black holes do not exist \cite{Emparan}, so we consider this point and
proceed with our work for the AdS case (i.e., $\Lambda $ is always negative
in this work).

One of the important differences between the obtained accelerating BTZ black
hole and the well-known BTZ black hole is related to the fact that the
asymptotical behavior of the accelerating BTZ black hole completely depends
on the acceleration parameter.

\begin{figure}[htbp]
\centering
\includegraphics[width=0.35\linewidth]{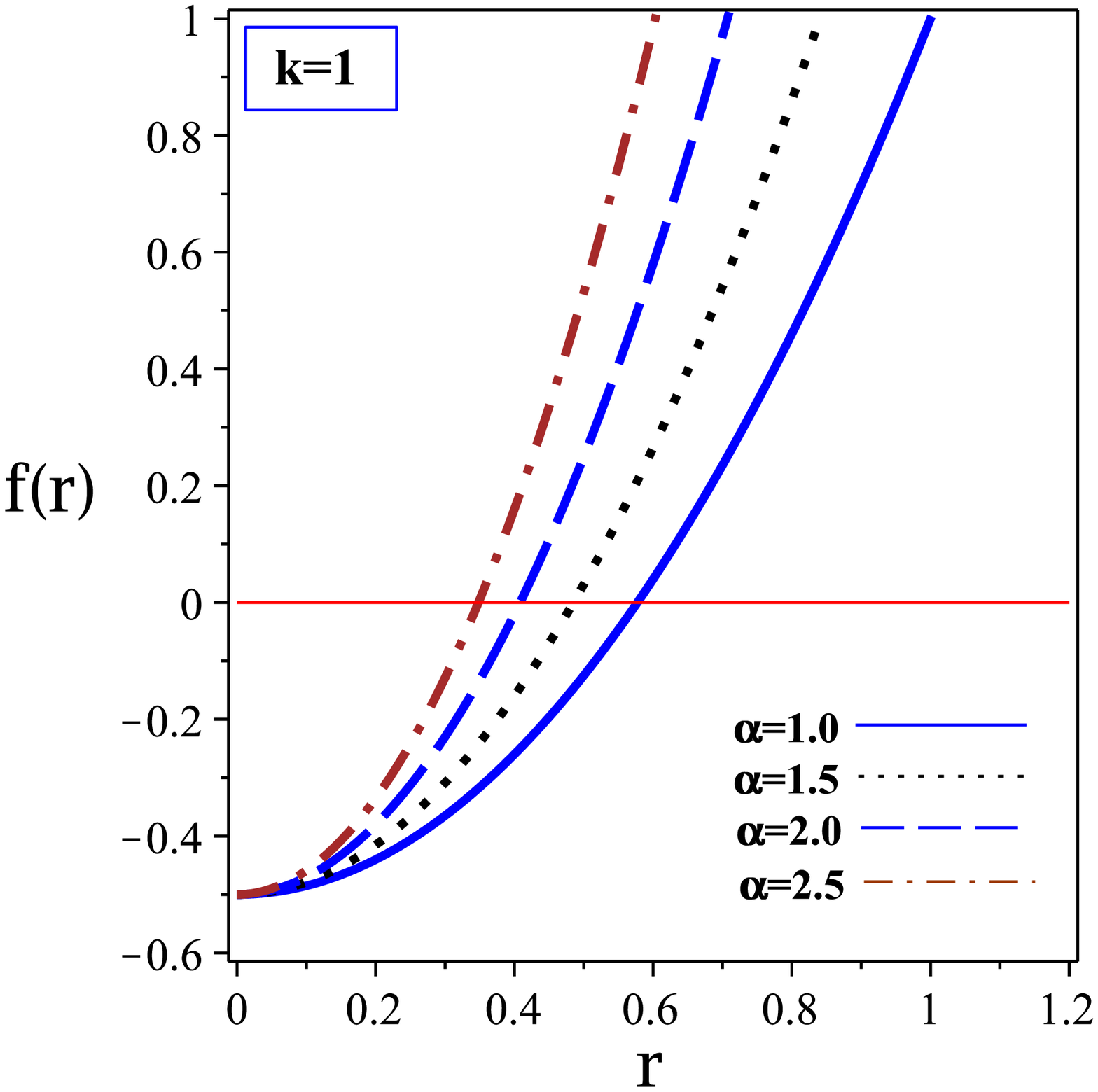} %
\includegraphics[width=0.35\linewidth]{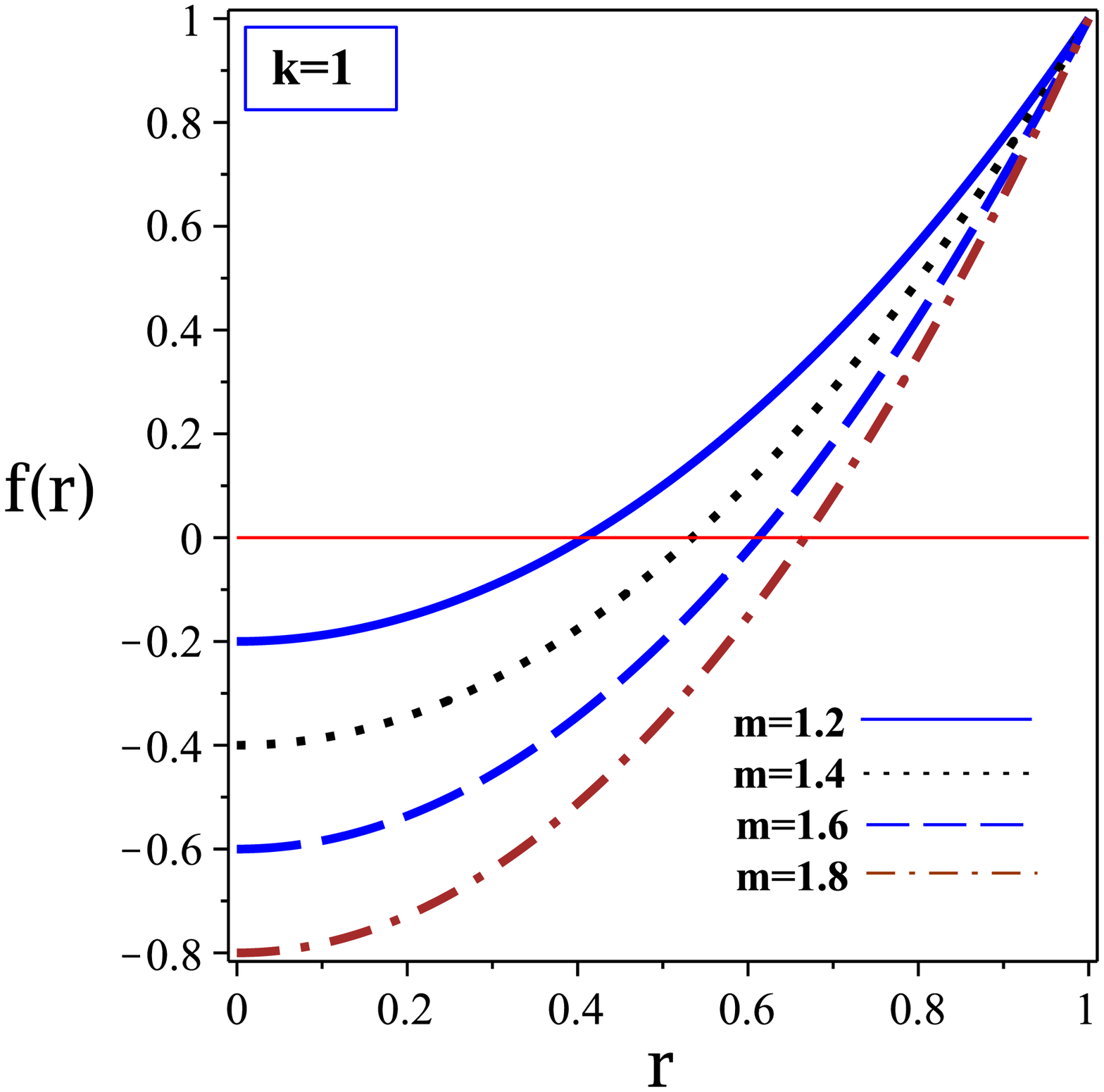} \newline
\includegraphics[width=0.35\linewidth]{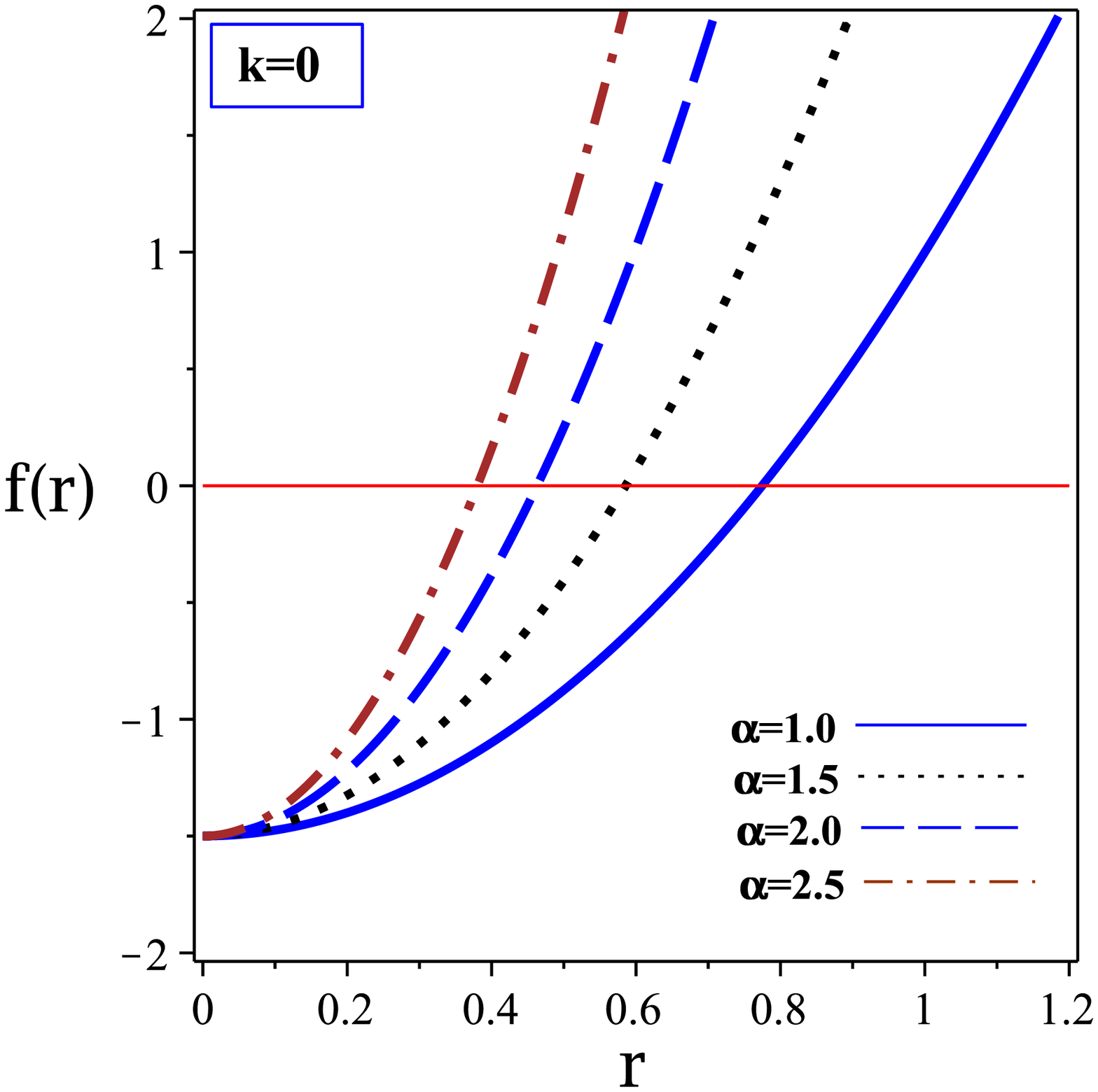} %
\includegraphics[width=0.35\linewidth]{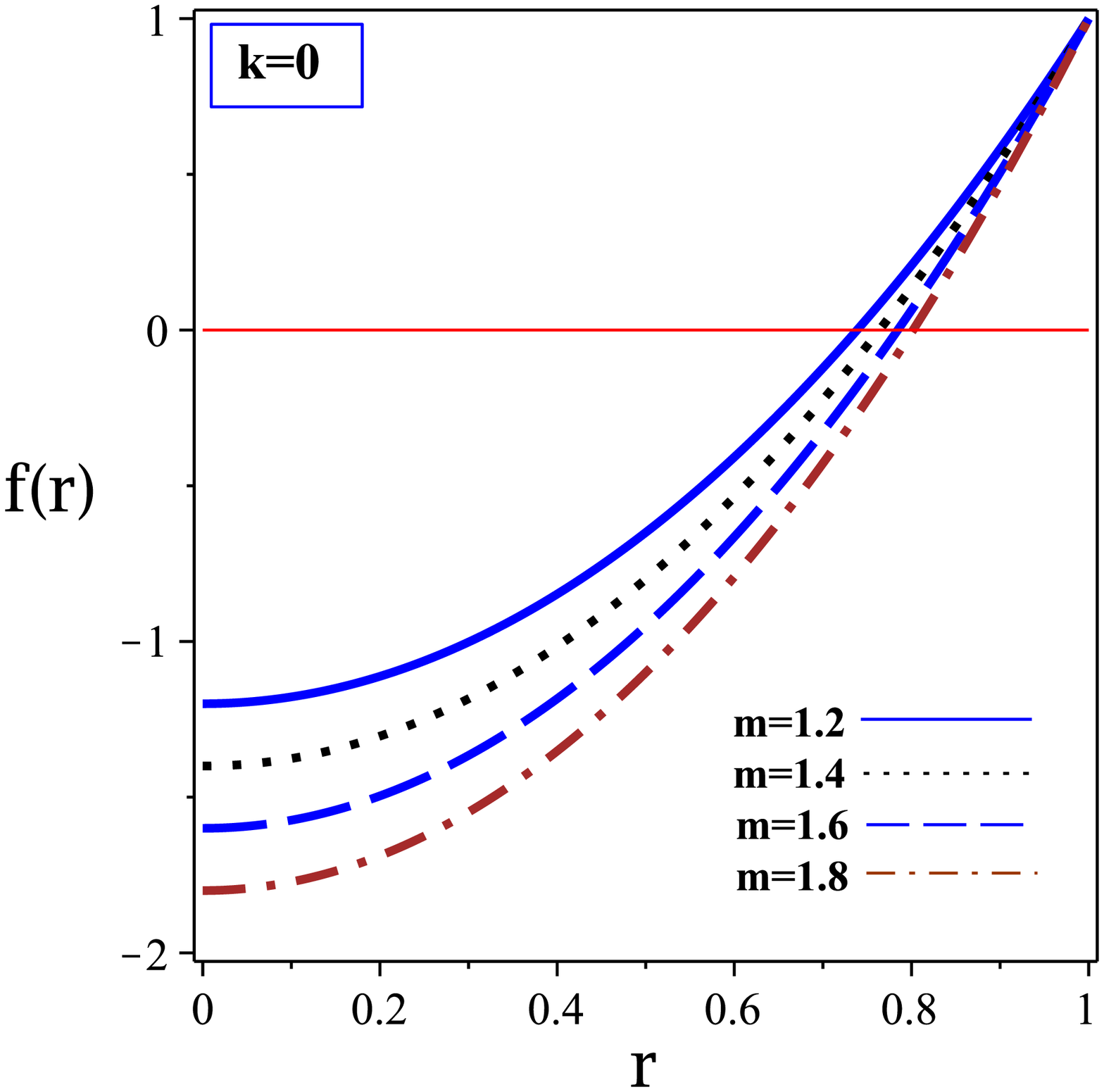} \newline
\includegraphics[width=0.35\linewidth]{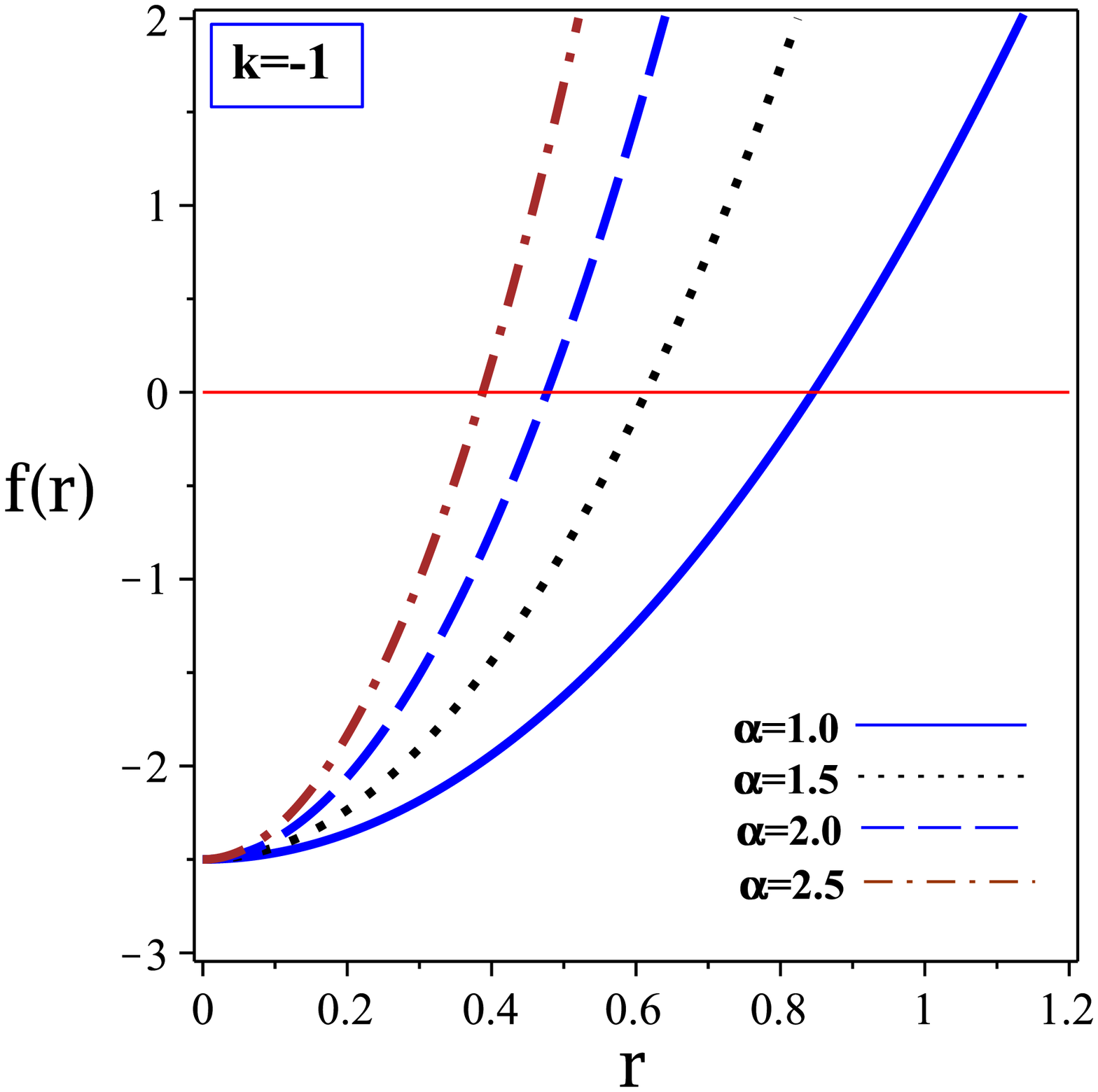} %
\includegraphics[width=0.35\linewidth]{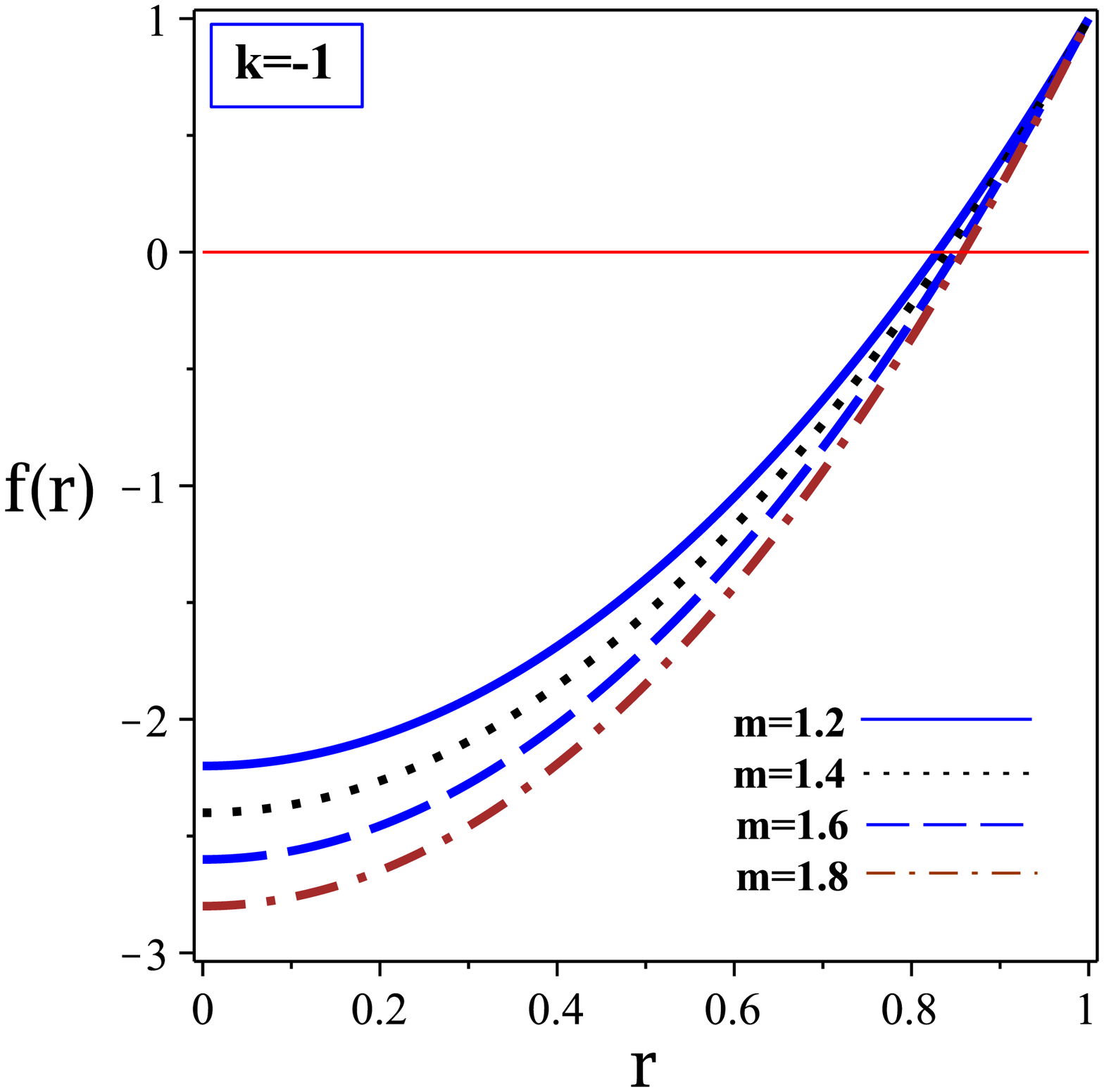} \newline
\caption{The metric function $f(r)$ versus $r$ for $\Lambda=-1$, and
different values of the topological parameter. Left panels for $m=1.5$.
Right panels for $\protect\alpha=1$.}
\label{Fig1}
\end{figure}

To study the obtained accelerating BTZ black holes, we plot the metric function (\ref{f(r)C1}) versus $r$ in Fig. \ref{Fig1}. According to the equation (\ref{rootuch}) and our results in Fig. \ref{Fig1}, the $(2+1)-$dimensional black holes with low acceleration have large radii. Indeed, by decreasing the acceleration parameter, the black holes have large radii (see the left panel in Fig. \ref{Fig1}). Also, by considering the negative values of the cosmological constant, massive black holes have large radii (see the right panel in Fig. \ref{Fig1}). Our results reveal that large black holes belong to $k=-1$ and small black holes have $k=1$ when other quantities have the same values. In other words, by considering the same values of parameters, there is a relation as $r_{+_{k=-1}}>r_{+_{k=0}}>r_{+_{k=1}}$ (see Fig. \ref{Fig1}, for more details).

\begin{figure}[htbp]
\centering
\includegraphics[width=0.37\linewidth]{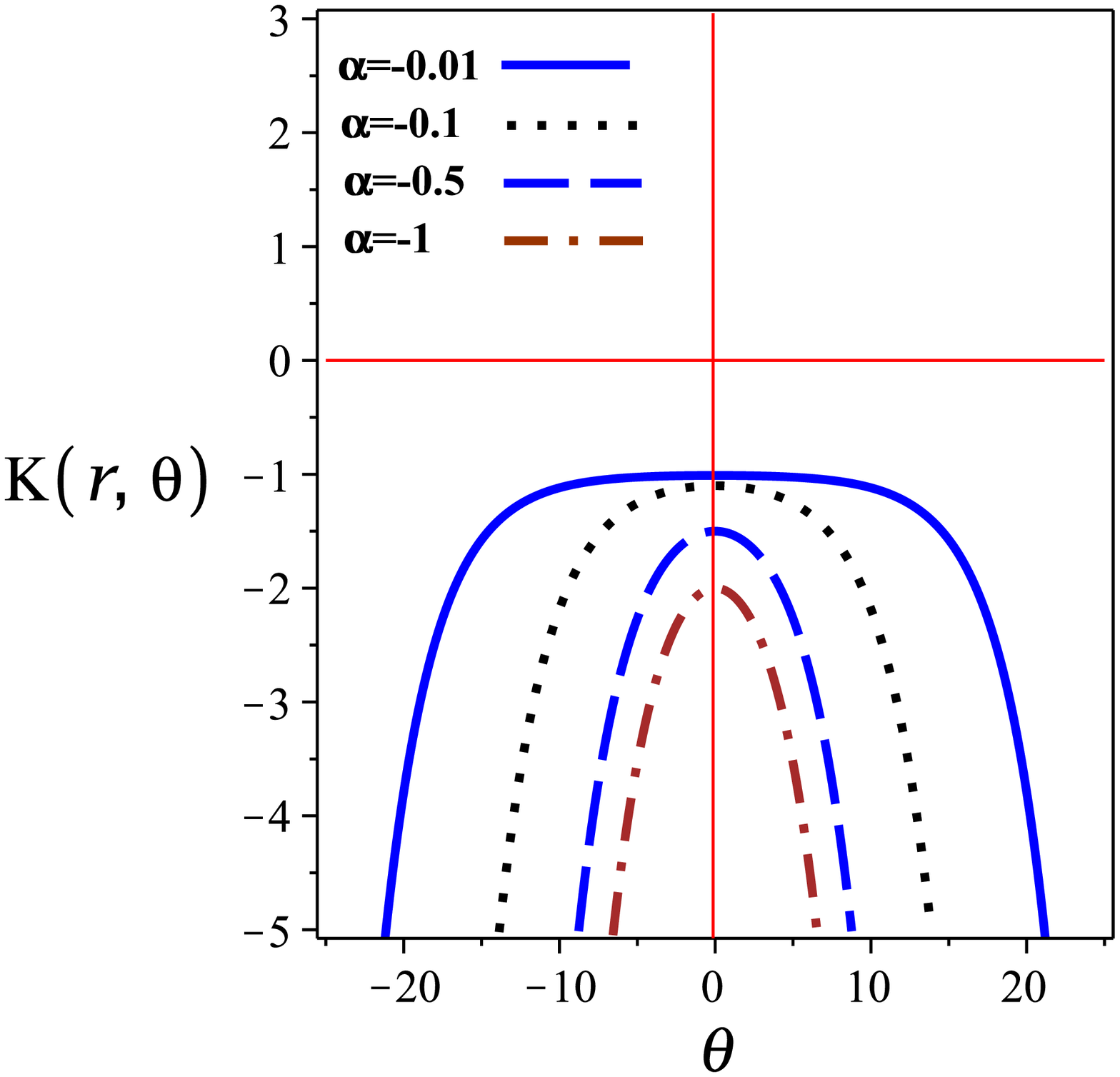} \includegraphics[width=0.37%
\linewidth]{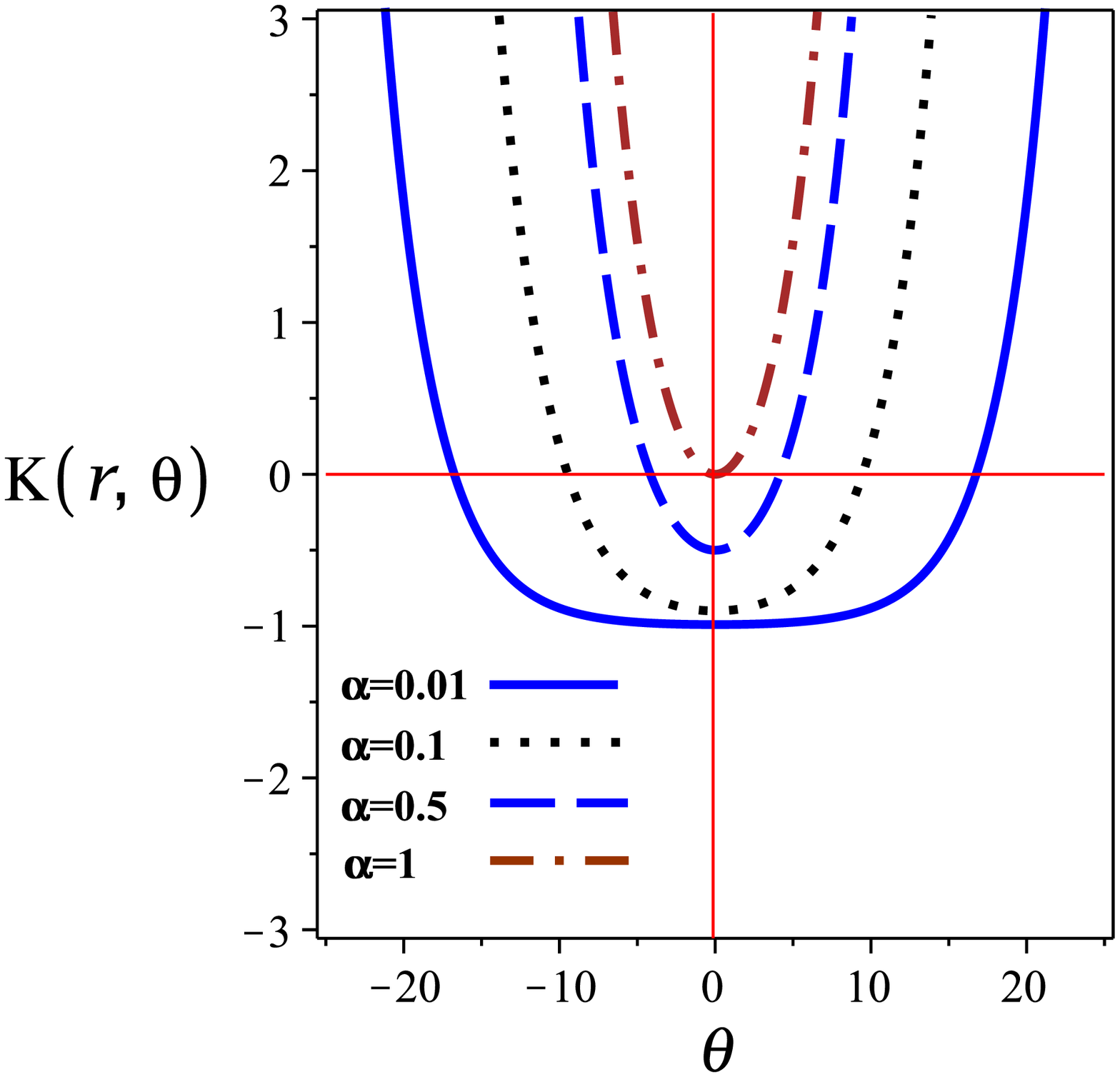}
\caption{$\mathcal{K}$$\left( r,\protect\theta \right)$ versus $\protect%
\theta$ for $k=1$, $m=1.1$ and $r=1$. Left panel for $\protect\alpha <0$ (or
(-) patch i.e., $\mathcal{K}\left( r,\protect\theta \right) < 0$). Right
panel for AdS case.}
\label{Fig1b}
\end{figure}

Now, we discuss the necessity of adjusting the range of $\theta $ in certain cases. According to Eq. (\ref{Metric}), every zero of $K\left( r,\theta\right) $ represents a conformal infinity in the metric. Considering $m>k$ and for $r>0$, the conformal factor will never have a zero when $\alpha <0$ (see the left panel in Fig. \ref{Fig1b}). Therefore, there is no worry about conformal infinities, but for $\alpha >0$, the conformal infinity exists. It is worthwhile to mention that the conformal infinity can intersect with the horizon when $0<\alpha r_{+}\leq 1$ (i.e., for AdS or flat cases), see the right panel in Fig. \ref{Fig1b}. This can be seen from the fact that the equation $K\left( r_{+},\theta \right) =0$ has two solutions $\theta =\pm \theta _{0}$, with the following form 
\begin{equation}
\theta _{0}=\frac{\arccos h\left( \frac{1}{\alpha r_{+}}\right) }{\sqrt{m-k}}%
,
\end{equation}%
the intersection appears when $\theta _{0}\in \left[ 0,\pi \right] $. It is notable that from the outside static observer's point of view, this means that the horizon stretches all the way through infinity. Indeed, it is non-compact. While evaluating the area of such horizons, one should exclude the part of the $r=r_{+}$ hypersurface that is hidden beyond the conformal
infinity.

The conformal infinity separates the spacetime into two patches: 
\begin{equation}
\left\{ 
\begin{array}{ccc}
\mathcal{K}\left( r,\theta \right) \leq 0 &  & (-)\text{ patch} \\ 
&  &  \\ 
\mathcal{K}\left( r,\theta \right) \geq 0 &  & (+)\text{ patch}%
\end{array}%
\right. ,
\end{equation}%
also, each observer can perceive only one of the two patches. So,
determination of the correct range for $\theta $ depends on which patch the
observer lives in. For the AdS case with $\theta _{0}\in \left[ 0,\pi \right]
$, we need to choose $\theta \in \left[ -\theta _{0},\theta _{0}\right] $ in
the ($-$) patch and $\theta \in \left[ -\pi ,\theta _{0})\cup (\theta
_{0},\pi \right] $ in the ($+$) patch (see the right panel in Fig. \ref%
{Fig1b}). For the flat case, it corresponds to $\theta _{0}=0$, horizon
intersects with the conformal infinity at a single point $\theta =0$ in the (%
$+$) patch, in which case we need to choose $\theta \in \left[ -\pi ,0)\cup
(0,\pi \right] $ (see the dotted-dashed line in the right panel from Fig. %
\ref{Fig1b}), and there is no horizon in the ($-$) patch (see Ref. \cite{Xu1}%
, for more details on the correct range of $\theta $).

In the following, we study the temperature of these black holes.

\subsection{Temperature}

In order to calculate the Hawking temperature, we employ the definition of surface gravity as
\begin{equation}
T_{H}=\frac{\kappa }{2\pi },  \label{Temp}
\end{equation}%
where $\kappa $\ is the surface gravity and is defined in the form $\kappa =%
\sqrt{\frac{-1}{2}\left( \nabla _{\mu }\chi _{\nu }\right) \left( \nabla
^{\mu }\chi ^{\nu }\right) }$.\ It is notable that $\chi =\partial _{t}$\ is
the Killing vector. Considering the metric (\ref{Metric}), the surface
gravity is given
\begin{equation}
\kappa =\left. \sqrt{\frac{\mathcal{H}_{1}^{2}\left( \mathcal{K}\left(
r,\theta \right) +1\right) ^{2}-2f^{\prime }\left( r\right) \mathcal{H}%
_{1}\left( \mathcal{K}\left( r,\theta \right) +1\right) -4\alpha ^{2}\left(
k-m\right) f\left( r\right) \sinh ^{2}\left( \sqrt{m-k}\theta \right)
+f^{\prime ^{2}}\left( r\right) }{4\mathcal{K}^{2}\left( r,\theta \right) }}%
\right\vert _{r=r_{+}},  \label{k1}
\end{equation}%
where $H_{1}=rf^{\prime }\left( r\right) -2f\left( r\right) $. The surface
gravity reduces to $\kappa =\frac{f^{\prime }\left( r\right) }{2}$, in the
absence of acceleration's parameter.

To extract the Hawking temperature of the uncharged accelerating BTZ black holes (Eq. (\ref{f(r)C1})), it is necessary to express the mass ($m$) in terms of the radius of event horizon $r_{+}$, the cosmological constant ($\Lambda $), the topological constant and the acceleration parameter $\alpha $. Equating $f\left( r\right) =0$, we can find $m$\ in the following form 
\begin{equation}
m=\frac{\left( k\alpha ^{2}+\Lambda \right) r_{+}^{2}-k}{\alpha
^{2}r_{+}^{2}-1},  \label{mm}
\end{equation}%
using the obtained metric function (\ref{f(r)C1}), and by substituting the mass (\ref{mm})\ within the equation (\ref{k1}), one can calculate the superficial gravity as 
\begin{equation}
\kappa \mathbf{=}\frac{\Lambda r_{+}}{\alpha ^{2}r_{+}^{2}-1}\mathbf{.}
\end{equation}%
where $r_{+}$ is the radius of the event horizon (or positive root in Eq. (\ref{rootuch})). Replacing the obtained metric function (\ref{f(r)C1}) into the relation (\ref{Temp}), we can obtain the temperature of the uncharged accelerating BTZ black holes as
\begin{equation}
T_{H}=\frac{\Lambda r_{+}}{2\pi \left( \alpha ^{2}r_{+}^{2}-1\right) }.
\label{TH1}
\end{equation}

Our result in Eq. (\ref{TH1}) indicates that the temperature depends on the cosmological constant and the acceleration parameter. It is noteworthy that there is a divergence point for the temperature at $r_{+}=\frac{1}{\alpha}$, and this point only depends on the acceleration parameter. On the other hand, we are interested in considering AdS spacetime, so we must respect $\alpha^{2}r_{+}^2<1$, which leads to $r_{+}<\frac{1}{\alpha }$. Therefore, there is no divergence point for the temperature. According to Eq. (\ref{TH1}), and applying the AdS case, the temperature is always positive. In the context of black holes, it is argued that the root of temperature ($T=0$) represents a border line between physical ($T>0$) and non-physical ($T<0$) black holes \cite{PanahPLB}. To find the physical black holes, we must obtain the temperature's root. It is clear that there is no root for the temperature, and for $\Lambda <0$, the temperature is always positive, and we encounter the physical case.

\section{Charged Accelerating BTZ Black Holes}

In this section, we want to extract accelerating BTZ black holes in the
presence of the electromagnetic field and the cosmological constant. We
first evaluate accelerating BTZ black holes in Einstein-Maxwell-$\Lambda $
gravity. Our findings indicate that by considering the Maxwell field, the
charged accelerating BTZ black hole solutions cannot recover the known
charged BTZ black holes (see Appendix A, for more details). For this
purpose, we extend the $(2+1)-$dimensional action (\ref{Action1}) in the
presence of the PMI NED field, which can be written as 
\begin{equation}
\mathcal{I}(g_{\mu \nu },A_{\mu })=\frac{1}{16\pi }\int_{\partial \mathcal{M}%
}d^{3}x\sqrt{-g}\left[ R-2\Lambda +L\left( \mathcal{F}\right) \right] ,
\label{actionPMI}
\end{equation}%
where $L\left( \mathcal{F}\right) =\left( -\mathcal{F}\right) ^{s}$, and $%
\mathcal{F}$ is the Maxwell invariant which is equal to $F_{\mu \nu }F^{\mu
\nu }$. It is noteworthy that $F_{\mu \nu }=\partial _{\mu }A_{\nu
}-\partial _{\nu }A_{\mu }$ is the electromagnetic tensor field, and $A_{\mu
}$ is the gauge potential. Also, $s$ is called the power of PMI NED.

To find the charged accelerating BTZ black hole solutions, we use a NED
field. We consider PMI NED for extracting charged accelerating BTZ black
hole solution in line with our purpose. In other words, we suppose $s=d/4$
in PMI NED (where $d$ is related to spacetime dimensions), known as
conformally invariant Maxwell (CIM) electrodynamics. According to this fact
that we want to investigate the $(2+1)-$dimensional black hole ($d=3$), we
have to consider $s=3/4$. So, $L\left( \mathcal{F}\right) $ in the action (%
\ref{actionPMI}) turns to $-\mathcal{F}^{3/4}$ (i.e., $L\left( \mathcal{F}%
\right) =-\mathcal{F}^{3/4}$). Varying the action (\ref{actionPMI}), with
respect to the gravitational field $g_{\mu \nu }$, and the gauge field $%
A_{\mu }$, yields 
\begin{eqnarray}
G_{\mu \nu }+\Lambda g_{\mu \nu } &=&T_{\mu \nu },  \label{Eqch1} \\
&&  \notag \\
\partial _{\mu }\left( \sqrt{-g}\mathcal{F}^{\frac{-1}{4}}F^{\mu \nu
}\right) &=&0,  \label{Eqch2}
\end{eqnarray}%
where $T_{\mu \nu }$is energy-momentum tensor. In the presence of CIM NED
for the $(2+1)$-dimensional spacetime, it can be written as 
\begin{equation}
T_{\mu \nu }=\frac{3}{2}F_{\mu \rho }F_{\nu }^{\rho }\left( -\mathcal{F}%
\right) ^{-\frac{1}{4}}+\frac{1}{2}g_{\mu \nu }\left( -\mathcal{F}\right) ^{%
\frac{3}{4}}.  \label{Tmunu}
\end{equation}

To obtain charged accelerating BTZ black holes, we consider a $(2+1)$%
-dimensional spacetime which is introduced in Eq. (\ref{Metric}). Since we
look for black hole solutions with a radial electric field, so the gauge
potential is given by $A_{\mu }=h(r)\delta _{\mu }^{0}$. Considering the
metric (\ref{Metric}), one can show that Eq. (\ref{Eqch2}) reduces to 
\begin{equation}
rh^{\prime \prime }\left( r\right) +2h^{\prime }\left( r\right) =0,
\label{Eqh(r)}
\end{equation}%
solving Eq. (\ref{Eqh(r)}), one can find $h(r)=\frac{q}{r}$, where $q$ is an
integration constant that is related to the electric charge, and the
electric field in the $(2+1)-$dimensional spacetime is given by $%
F_{tr}=-F_{rt}=\frac{q}{r^{2}}$, where it is inverse square of $r$.

Now, we should find a suitable metric function to satisfy all components of Eq. (\ref{Eqch1}). Considering the obtained $h\left( r\right) $, one can
show that Eq. (\ref{Eqch1}) reduces to 
\begin{eqnarray}
Eq_{tt} &=&Eq_{rr}=r^{2}\left[ \left( \frac{1-\alpha r}{2}\right) f^{\prime
}\left( r\right) +\alpha f\left( r\right) +r\left( \Lambda +\left(
k-m\right) \alpha ^{2}\right) +\alpha \left( m-k\right) \right] +\frac{%
\left( 2q^{2}\right) ^{3/4}\left( \alpha r-1\right) ^{3}}{4},
\label{Eqttch1} \\
&&  \notag \\
Eq_{\theta \theta } &=&\left( \alpha r-1\right) ^{2}f^{\prime \prime }\left(
r\right) -2\alpha \left( \alpha r-1\right) f^{\prime }\left( r\right)
+2\alpha ^{2}f\left( r\right) +2\Lambda -\frac{\left( 2q^{2}\right)
^{3/4}\left( \alpha r-1\right) ^{3}}{r^{3}}.  \label{Eqthetach1}
\end{eqnarray}

Notably, we supposed $\theta =0$ in Eq. (\ref{Eqch1}), similar to the uncharged case. Using the equations (\ref{Eqttch1}) and (\ref{Eqthetach1}), we are in a position to obtain the metric function $f\left( r\right) $. After some calculation, we find an exact general solution in the following form 
\begin{eqnarray}
f(r) &=&\alpha ^{2}r^{2}C\left( m,k,\Lambda ,\alpha ,q\right) +2\alpha
r\left( m-k-C\left( m,k,\Lambda ,\alpha ,q\right) \right) +2\left(
k-m\right) +C\left( m,k,\Lambda ,\alpha ,q\right)  \notag \\
&&  \notag \\
&&+\frac{\Lambda }{\alpha ^{2}}-\frac{2\Lambda r}{\alpha }-\frac{\left(
2q^{2}\right) ^{3/4}\left( \alpha r-1\right) ^{2}}{2r},  \label{f(r)Ch}
\end{eqnarray}%
where $C\left( m,k,\Lambda ,\alpha ,q\right) $ is a constant which depends $%
m $, $k$, $\Lambda $, $\alpha $, and $q$. We should note that the solution (%
\ref{f(r)Ch}), satisfy all field equations of Eq. (\ref{Eqch1}).

It is necessary that the obtained charged accelerating BTZ black hole
solution in Eq. (\ref{f(r)Ch}) reduces to known charged BTZ black hole in
the Einstein-CIM gravity \cite{CIM} when ignoring the acceleration
parameter. For this purpose, we consider $C\left( m,k,\Lambda ,\alpha
,q\right) $ in the following form 
\begin{equation}
C\left( m,k,\Lambda ,\alpha ,q\right) =m-k-\frac{\Lambda }{\alpha ^{2}}-%
\frac{\left( 2q^{2}\right) ^{3/4}\alpha }{4},  \label{C2}
\end{equation}%
which the solution (\ref{f(r)Ch}) turns to%
\begin{equation}
f\left( r\right) =k-m+\left( \left( m-k\right) \alpha ^{2}-\Lambda \right)
r^{2}-\frac{\left( 2q^{2}\right) ^{\frac{3}{4}}\left( \alpha r+2\right)
\left( \alpha r-1\right) ^{2}}{4r}.  \label{f(r)Ch2}
\end{equation}

It is clear that in the absence of the acceleration parameter, the solution (\ref{f(r)Ch2}) reduces to the charged BTZ black hole in the Einstein-CIM
gravity \cite{CIM} 
\begin{equation}
f\left( r\right) _{BTZ-CIM}=k-m-\Lambda r^{2}-\frac{\left( 2q^{2}\right) ^{%
\frac{3}{4}}}{2r}.  \label{MBTZacc}
\end{equation}
where the above solution is an exact solution to the Einstein-CIM gravity
with cosmological constant $\Lambda $ when the acceleration parameter is
zero. Besides, the equation (\ref{f(r)Ch2}) turns to the obtained uncharged
accelerating BTZ black holes (Eq. (\ref{f(r)C1})) by considering $q=0$.

\subsection{Curvature Scalars}

In order to look for the singularity(ies) of spacetime for the obtained
solutions (\ref{f(r)Ch2}), we evaluate the Ricci and Kretschmann scalars.
Using the metric function (\ref{f(r)Ch2}) and the equations (\ref{RUnch1})
and (\ref{KUnCh1}), we obtain these scalars which are 
\begin{eqnarray}
R &=&6\Lambda ,  \notag \\
&& \\
R_{\alpha \beta \gamma \delta }R^{\alpha \beta \gamma \delta } &=&12\Lambda
^{2}+\frac{3\sqrt{2}q^{3}\left( \alpha r-1\right) ^{6}}{r^{6}},  \notag
\end{eqnarray}%
where the Kretschmann scalar indicates that there is a singularity at $r=0$
(i.e., $\underset{r\longrightarrow 0}{\lim }R_{\alpha \beta \gamma \delta
}R^{\alpha \beta \gamma \delta }\longrightarrow \infty $). It is notable
that the divergence point removes when $\alpha r=1$. In other words,
we can remove the singularity of Kretschmann scalar by considering $\alpha
r=1$.

According to the obtained solutions (\ref{f(r)Ch2}), the electrical charge
affects their asymptotical behavior. Indeed, the asymptotical behavior is
dependent on $\Lambda $, $\alpha $, $m$, $k$ and $q$. So, the obtained
spacetime for the charged case is not exactly asymptotically AdS due to the
existence of different parameters in the asymptotical behavior of spacetime.

\subsection{Horizon Structure}

To find the real roots of the obtained metric function (\ref{f(r)Ch2}), we
solve $\left. f(r)\right\vert _{r=r_{+}}=0$, which leads to 
\begin{equation}
\Lambda =\frac{-2\left( 2q^{2}\right) ^{\frac{3}{4}}\left( \alpha
r_{+}+2\right) \left( \alpha r_{+}-1\right) ^{2}+4r_{+}\left( m-k\right)
\left( \alpha ^{2}r_{+}^{2}-1\right) }{4r_{+}^{3}},
\end{equation}%
where imposes the following conditions to have AdS spacetime 
\begin{equation}
\Lambda <0\Rightarrow \left\{ 
\begin{array}{c}
\alpha r_{+}+2>0 \\ 
\\ 
\alpha ^{2}r_{+}^{2}-1<0%
\end{array}%
\right. ,
\end{equation}%
and by using the above condition we get the AdS limit in following form 
\begin{equation}
-2<\alpha r_{+}<1,
\end{equation}%
where we obtain the above limit by considering $m>k$.

To more study the effect of topological constant on the obtained charged
accelerating BTZ black holes, we consider $k=$ $\pm 1$ and $0$.

\subsubsection{\textbf{Case:}$\ k=1$}

\begin{figure}[htbp]
\centering
\includegraphics[width=0.31\linewidth]{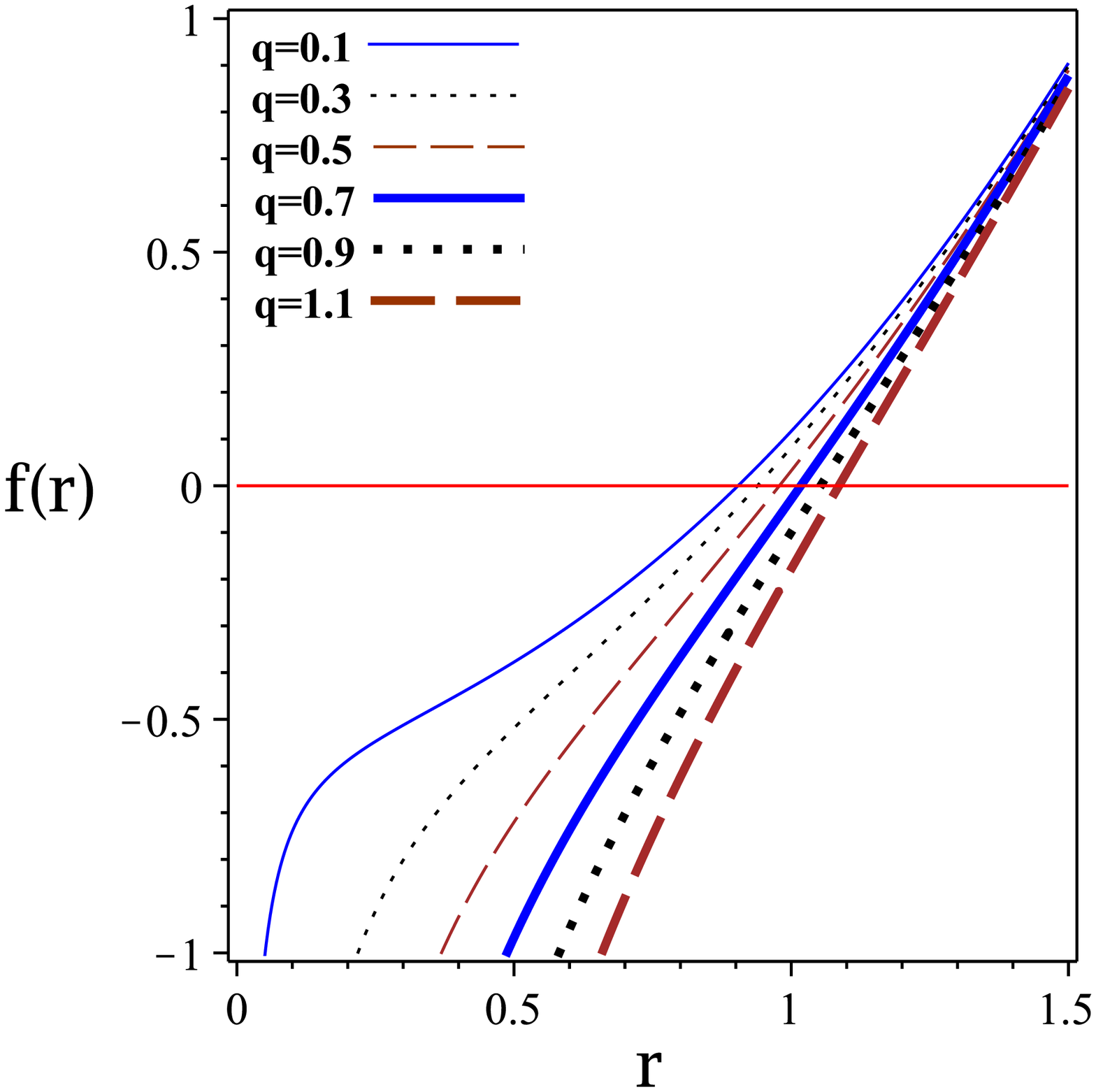} %
\includegraphics[width=0.31\linewidth]{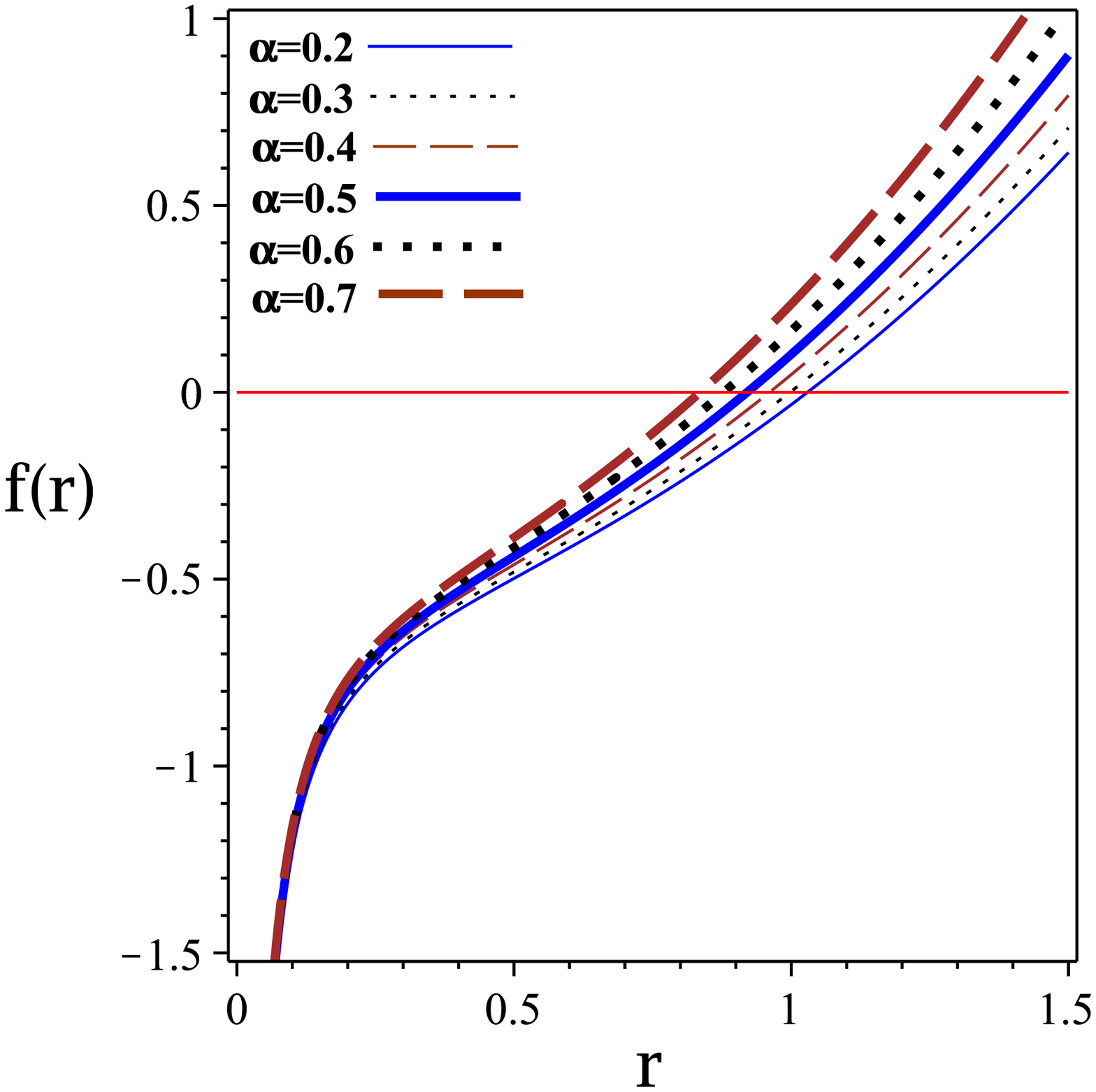} %
\includegraphics[width=0.31\linewidth]{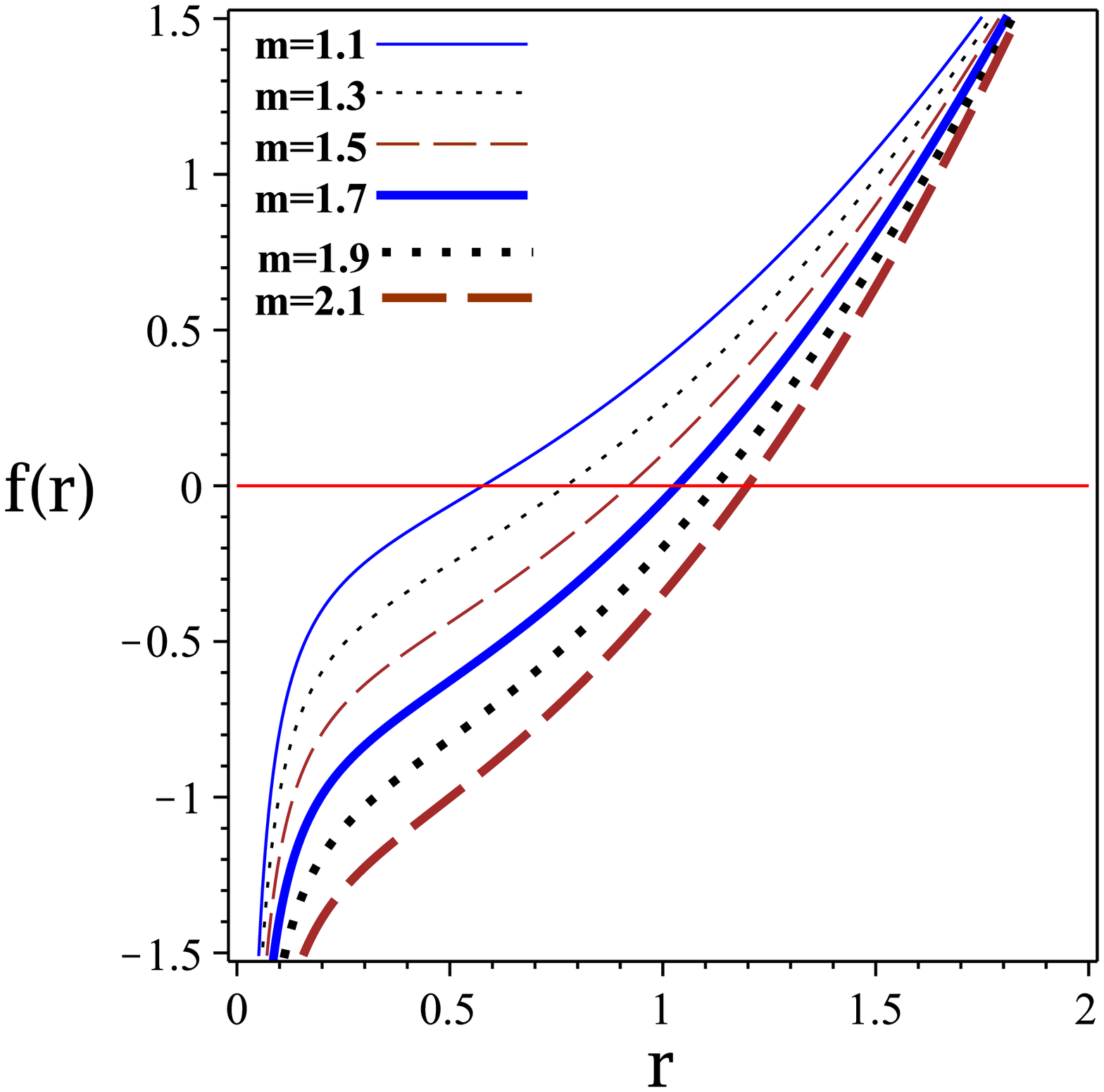}
\caption{The metric function $f(r)$ versus $r$ for $k=1$ and $\Lambda=-0.5$.
Left panel for $\protect\alpha=0.5$, and $m=1.5$. Middle panel for $q=0.2$,
and $m=1.5$. Right panel for $q=0.2$, and $\protect\alpha=0.5$.}
\label{Fig3}
\end{figure}

Our results indicate that the radius of black hole increases by increasing
the electrical charge and mass parameters. Indeed, massive black holes with
high electrical charge have large radii (see left and right panels in Fig. %
\ref{Fig3}). On the other hand, there is a different behavior for the
acceleration parameter. In other words, the charged BTZ black holes with
high acceleration have small radii (see the middle panel in Fig. \ref{Fig3}%
). As a result, the accelerating BTZ black holes with high mass and strong
electrical charges have large radii.

\subsubsection{\textbf{Case }$k=-1$}

\begin{figure}[tbph]
\centering
\includegraphics[width=0.31\linewidth]{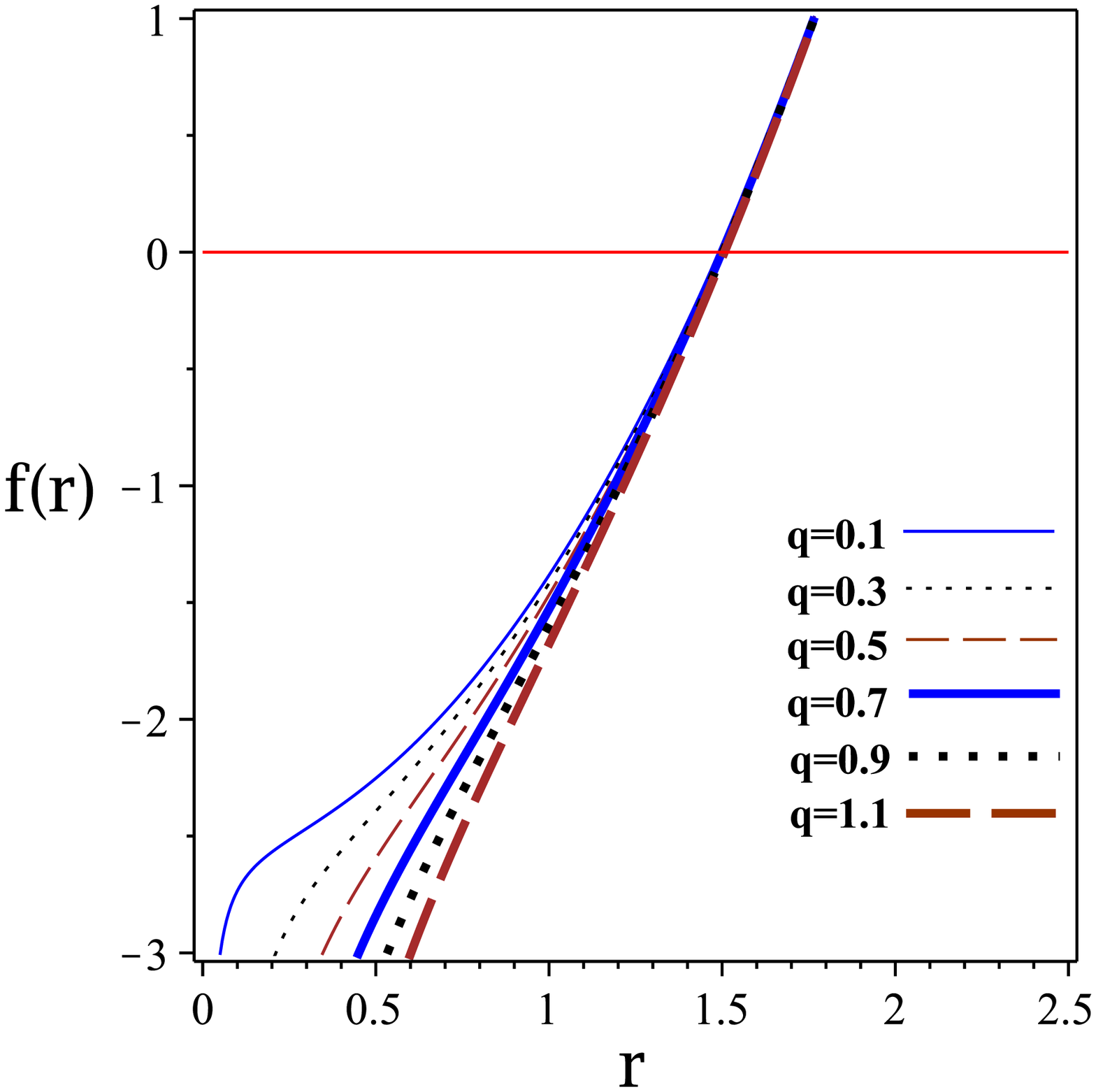} %
\includegraphics[width=0.31\linewidth]{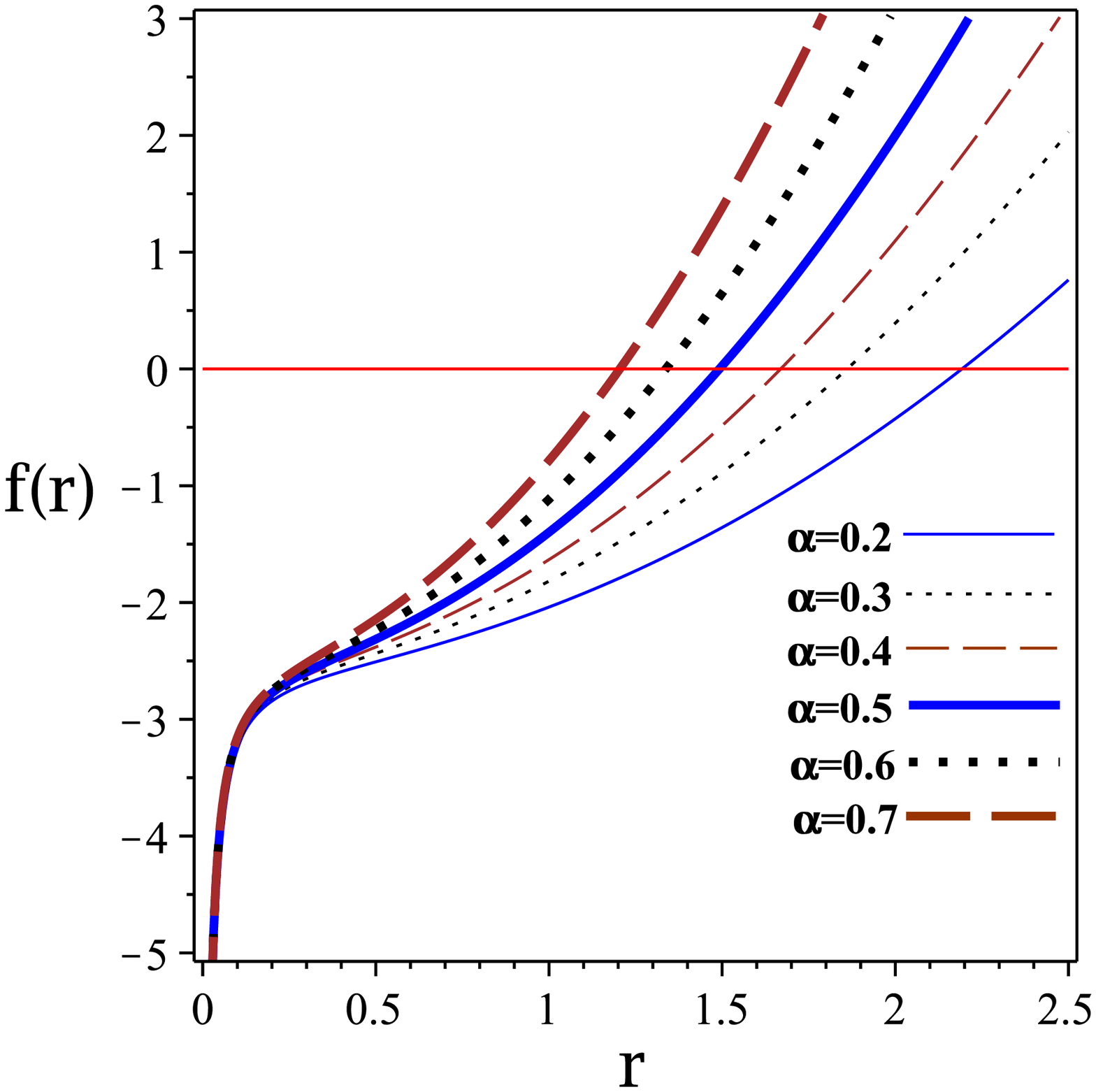} %
\includegraphics[width=0.31\linewidth]{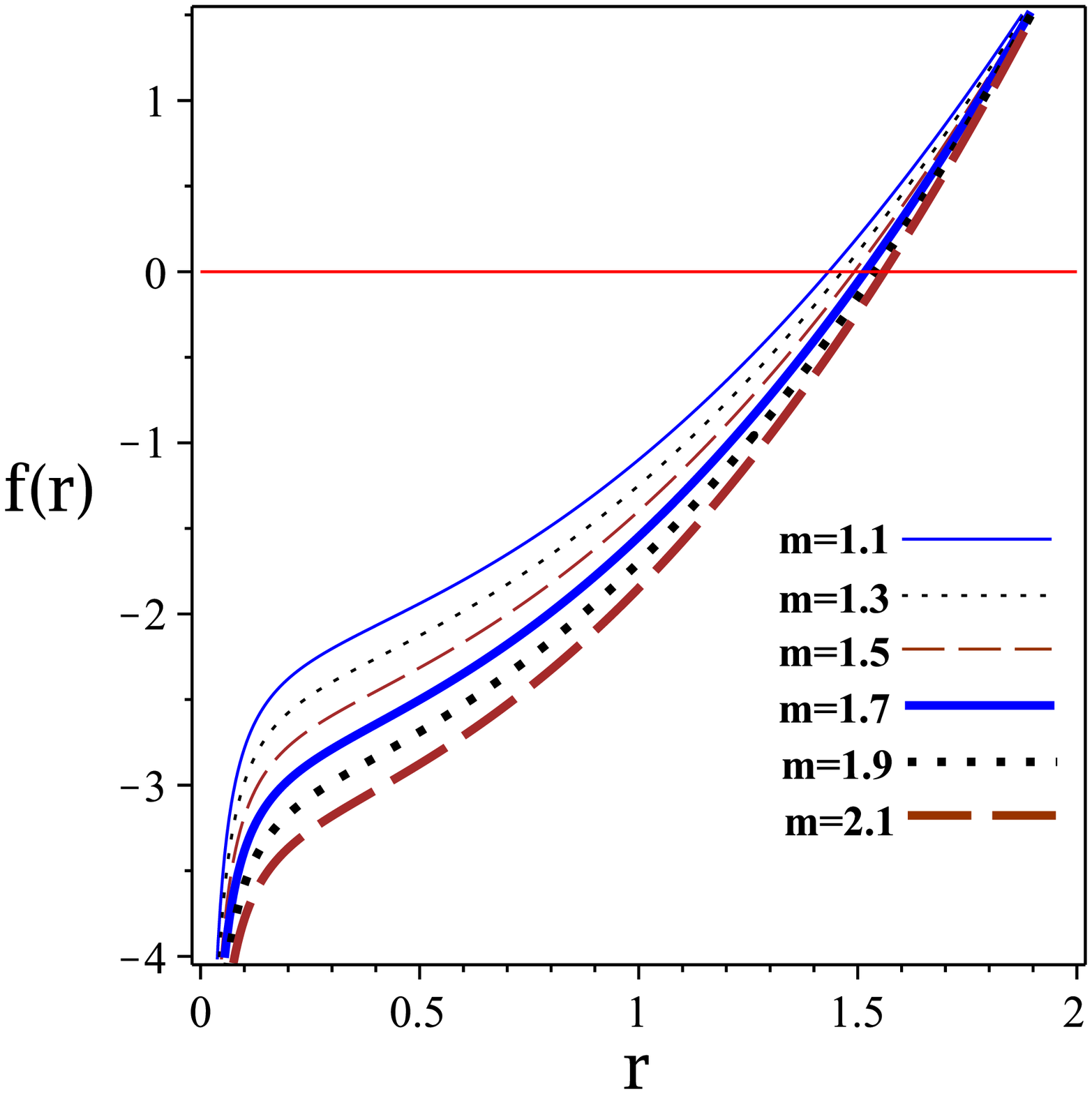}
\caption{The metric function $f(r)$ versus $r$ for $k=-1$, and $\Lambda=-0.5$%
. Left panel for $\protect\alpha =0.5$, and $m=1.5$. Middle panel for $q=0.2$%
, and $m=1.5$. Right panel for $q=0.2$, and $\protect\alpha =0.5$.}
\label{Fig3b}
\end{figure}

For this case, massive black holes with high electrical charge have large
radii (see left and right panels in Fig. \ref{Fig3b}). It means that the
horizon increases by increasing mass and the electrical charge. However, the
black holes with high acceleration have small radii (see the middle panel in
Fig. \ref{Fig3b}). As a result, the largest accelerating BTZ black holes
have large masses and strong electrical charges. Our analysis indicates that
the behavior of the horizon for $k=-1$ is similar to $k=1$, when $\Lambda <0$
(see Figs. \ref{Fig3} and \ref{Fig3b}). Also, comparing the obtained results
in Figs. \ref{Fig3} and \ref{Fig3b}, and by considering the same values for
parameters, we find that the black holes with $k=-1$ are larger than $k=1$.

\subsubsection{Case $k=0$}

\begin{figure}[tbph]
\centering
\includegraphics[width=0.31\linewidth]{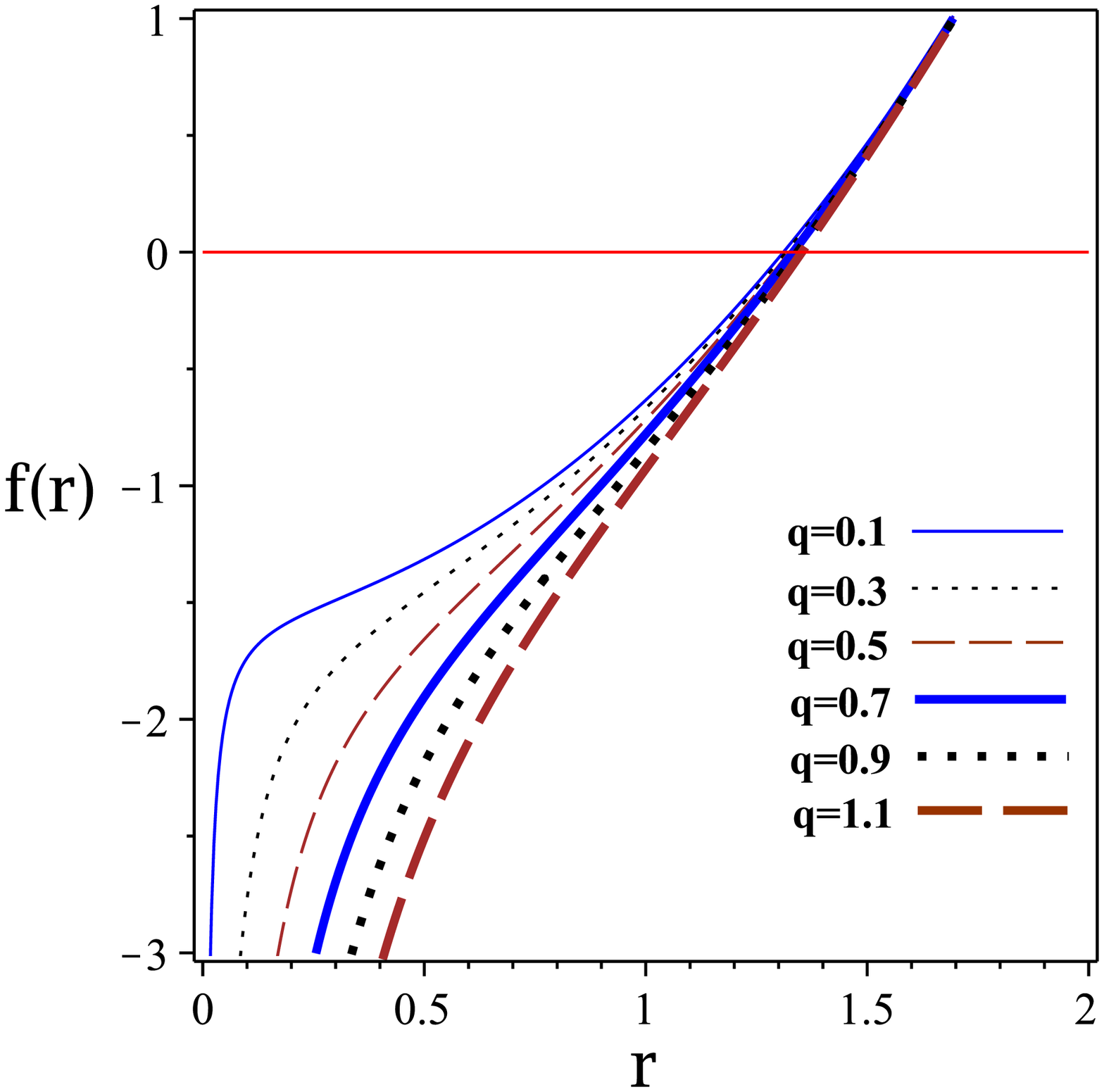} %
\includegraphics[width=0.31\linewidth]{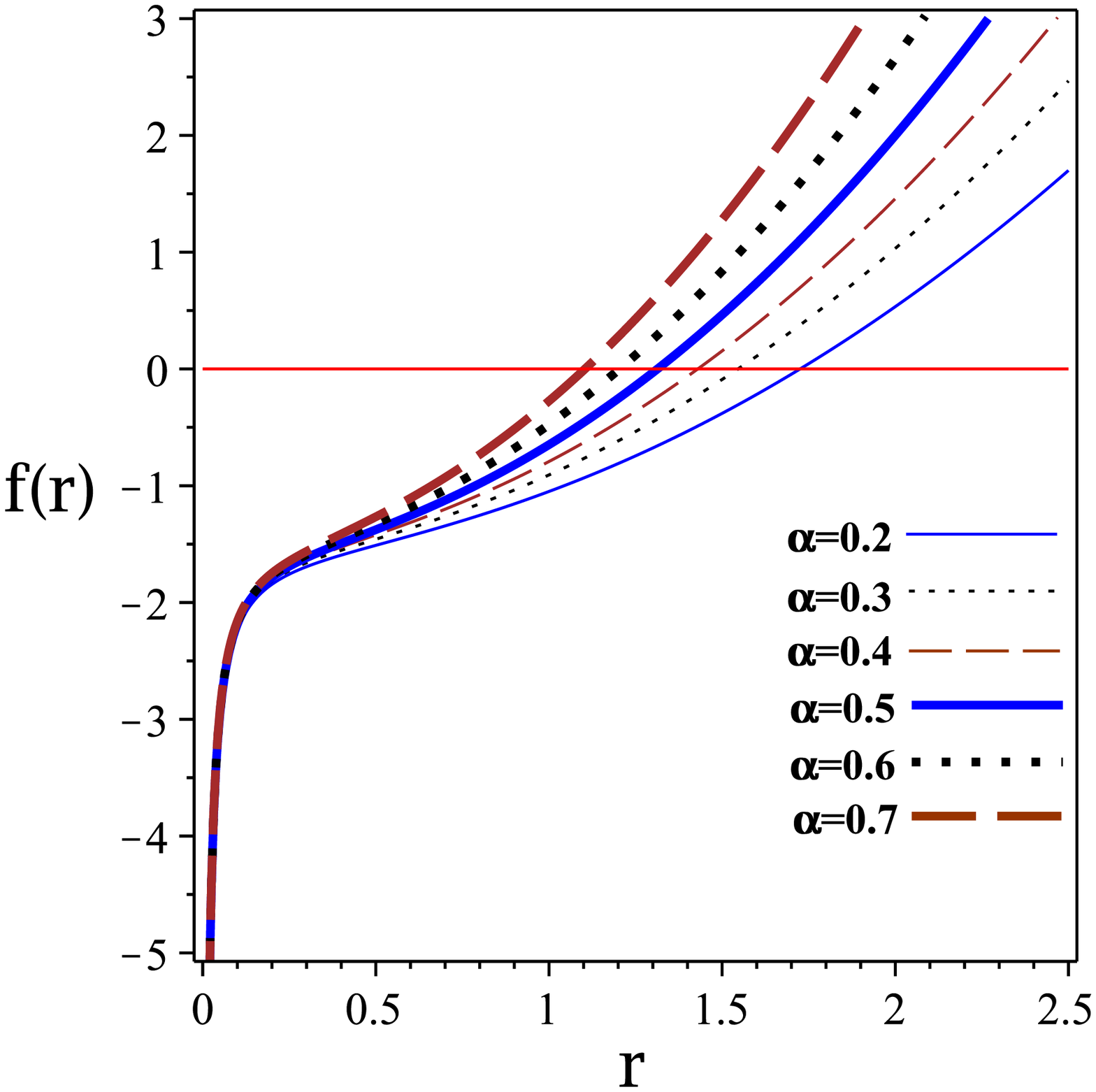} %
\includegraphics[width=0.31\linewidth]{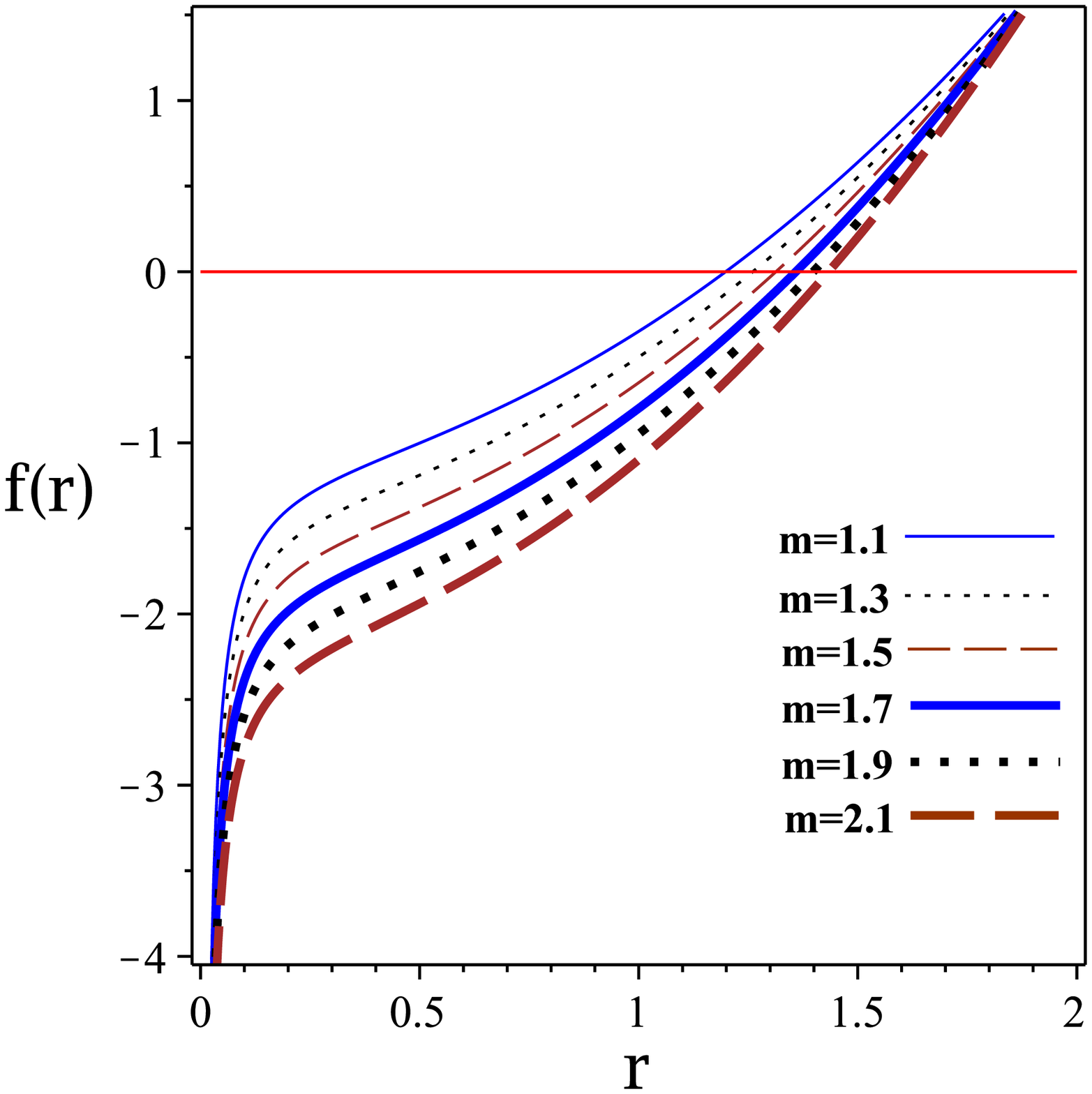}
\caption{The metric function $f(r)$ versus $r$ for $k=0$ and $\Lambda=-0.5$.
Left panel for $\protect\alpha =0.5$, and $m=1.5$. Middle panel for $q=0.2$,
and $m=1.5$. Right panel for $q=0.2$, and $\protect\alpha =0.5$.}
\label{Fig3c}
\end{figure}

There are the same behaviors similar to previous cases. In other words,
massive black holes with large values of electrical charge are big black
holes (see left and right panels in Fig. \ref{Fig3c}). Also, the radius of
black holes decreases by increasing the acceleration parameter. In this
case, large black holes have small acceleration (see the middle panel in
Fig. \ref{Fig3c}).

We evaluate the effects of the electrical charge, the acceleration
parameter, and mass on the horizon of charged accelerating BTZ black holes
with different topological constants in Table. \ref{tab1}. Our analysis
indicates that black holes with topological constant $k=-1$ are bigger than $%
k=0$, and $k=1$. In other words, by considering the same values of
parameters, black holes have large radii with topological constant $k=-1$ in
comparison with other values of the topological constant (i.e., $k=0$, and $%
k=1$). In addition, the smallest black holes have topological constant $k=1$.

\begin{table*}[tbp]
\caption{The radius of black hole with different values of topological
constant ($k$) for $\Lambda =-0.5$. The first table for different $q$ with $%
\protect\alpha=0.5$ and $m=1.5$. The second table for different $\protect%
\alpha$ with $q=0.2$ and $m=1.5$. The third table for different $m$ with $%
\protect\alpha =0.5$ and $q=0.2$.}
\label{tab1}
\begin{center}
\begin{tabular}{|c|c|c|c|c|}
\hline
$k$ & $q=0.1$ & $q=0.3$ & $q=0.5$ & $q=0.7$ \\ \hline\hline
$+1$ & $r_{+}=0.904$ & $r_{+}=0.939$ & $r_{+}=0.978$ & $r_{+}=1.017$ \\ 
\hline
$0$ & $r_{+}=1.310$ & $r_{+}=1.316$ & $r_{+}=1.324$ & $r_{+}=1.333$ \\ \hline
$-1$ & $r_{+}=1.491$ & $r_{+}=1.493$ & $r_{+}=1.496$ & $r_{+}=1.499$ \\ 
\hline
\end{tabular}%
\par
\begin{tabular}{|c|c|c|c|c|}
\hline
$k$ & $\alpha=0.1$ & $\alpha=0.3$ & $\alpha=0.5$ & $\alpha=0.7$ \\ 
\hline\hline
$+1$ & $r_{+}=1.053$ & $r_{+}=0.997$ & $r_{+}=0.920$ & $r_{+}=0.835$ \\ 
\hline
$0$ & $r_{+}=1.725$ & $r_{+}=1.545$ & $r_{+}=1.313$ & $r_{+}=1.103$ \\ \hline
$-1$ & $r_{+}=2.192$ & $r_{+}=1.861$ & $r_{+}=1.492$ & $r_{+}=1.204$ \\ 
\hline
\end{tabular}%
\par
\begin{tabular}{|c|c|c|c|c|}
\hline
$k$ & $m=1.1$ & $m=1.3$ & $m=1.5$ & $m=1.7$ \\ \hline\hline
$+1$ & $r_{+}=0.578$ & $r_{+}=0.773$ & $r_{+}=0.920$ & $r_{+}=1.033$ \\ 
\hline
$0$ & $r_{+}=1.198$ & $r_{+}=1.260$ & $r_{+}=1.313$ & $r_{+}=1.358$ \\ \hline
$-1$ & $r_{+}=1.433$ & $r_{+}=1.464$ & $r_{+}=1.492$ & $r_{+}=1.517$ \\ 
\hline
\end{tabular}%
\end{center}
\end{table*}

Another interesting quantity that gives us information on the obtained
charged accelerating BTZ black holes is related to the Hawking temperature.

\subsection{Temperature}

To extract the Hawking temperature of the charged accelerating BTZ black holes (\ref{f(r)Ch2}), we express the mass ($m$) in terms of the radius of event horizon $r_{+}$, the cosmological constant $\Lambda $, the topological constant, the acceleration parameter $\alpha $, and the electrical charge $q$, by equating $f\left( r\right) =0$, which leads to
\begin{equation}
m=\frac{\left( k\alpha ^{2}+\Lambda \right) r_{+}^{2}-k}{\alpha
^{2}r_{+}^{2}-1}+\frac{\left( 2q^{2}\right) ^{3/4}\left( \alpha
r_{+}+2\right) \left( \alpha r_{+}-1\right) }{8\left( \alpha r_{+}+1\right)
r_{+}},  \label{mmch}
\end{equation}%
using the obtained metric function (\ref{f(r)Ch2}), and by substituting the mass (\ref{mmch})\ within the equations (\ref{k1}) and (\ref{Temp}), we can calculate the temperature of charged accelerating BTZ black holes in the following form 
\begin{equation}
T_{H}=\frac{\Lambda r_{+}}{2\pi \left( \alpha ^{2}r_{+}^{2}-1\right) }-\frac{%
\left( 2q^{2}\right) ^{3/4}\left( 2\alpha r_{+}+1\right) \left( \alpha
r_{+}-1\right) ^{2}}{8\pi \left( \alpha ^{2}r_{+}^{2}-1\right) r_{+}^{2}},
\label{THCh1}
\end{equation}%
where $r_{+}$ is the radius of the black hole. The temperature of these black holes depends on the cosmological constant, the acceleration parameter, and the electrical charge. It is notable that in the absence of the electrical charge Eq. (\ref{THCh1}) turns to Eq. (\ref{TH1}).

Here, we evaluate temperature behavior. Our analysis indicates that, there is no root for the temperature when $\Lambda <0$. However, there is a divergence point of temperature at $\alpha ^{2}r_{+}^{2}=1$, but we cannot consider it because we have to respect the AdS limit (i.e., $-2<\alpha r_{+}<1$). Therefore, we do not have any divergence point. By looking at the equation (\ref{THCh1}), it is clear that for the negative value of the cosmological constant, the temperature is always positive. In addition, there is an interesting effect of $\alpha $ on the temperature. Applying the AdS limit ($-2<\alpha r_{+}<1$), the positive area of the temperature decreases by increasing the value of the acceleration parameter. In other words, charged BTZ AdS black holes with high acceleration have a small physical area.

\section{\noindent Closing Remarks}

In this paper, we obtained uncharged accelerating BTZ AdS black hole solutions. The asymptotical behavior of this spacetime was not exactly AdS due to different parameters. In order to respect the signature of spacetime, we extracted two conditions, which are $m>k$ and $\Lambda <\left( m-k\right)
\alpha ^{2}$. These equations revealed that the mass of black holes must be more than the topological constant. Our analysis of the behavior of the metric function indicated that massive BTZ black holes with low acceleration have large radii when $\Lambda <0$.

We obtained the temperature of these black holes. The temperature was dependent on the acceleration parameter and the cosmological constant. More studies indicated that the temperature was always positive, and the physical black holes were in the radii less than $\frac{1}{\alpha }$ ($r_{+}<\frac{1}{\alpha }$).

Then, we evaluated accelerating BTZ black holes in the presence of the Maxwell field. Our findings showed that charged accelerating BTZ black holes could not recover the known BTZ black holes without the acceleration parameter. This result motivated us to consider a NED field instead of a Maxwell field. Then, we obtained exact charged accelerating BTZ black hole solutions by coupling GR with CIM NED in the presence of the cosmological constant. The asymptotical behavior of spacetime was dependent on the cosmological constant, the electrical charge, mass, the acceleration parameter, and the topological constant, and it was not exactly asymptotically AdS.

We evaluated the metric function and found one real root. Then, we studied
the effects of charge and acceleration parameters on this root in Fig. \ref%
{Fig3}. Our findings indicated that; i) strongly charged accelerating BTZ
AdS black holes had large radii. ii) the charged accelerating BTZ AdS black
holes had small radii when $\alpha $ increased. iii) charged accelerating
BTZ AdS black holes with topological constant $k=-1$ were bigger than $k=0$,
and $k=1$. In other words, by considering the same values of parameters,
black holes had large radii with topological constant $k=-1$ in comparison
with other values of the topological constant (i.e., $k=0$, and $k=1$).
Also, the smallest black holes had topological constant $k=1$.

We extracted the temperature of the obtained charged accelerating BTZ black
holes. Our calculations indicated that it depended on $q$, $\alpha $, and $%
\Lambda $. Also, we found that the temperature was always positive when $%
\Lambda <0$, and the positive area of the temperature decreased by
increasing the value of the acceleration parameter.

\begin{acknowledgements}
I would like to thank Prof. Ruth Gregory for the helpful discussions. This work has been supported financially by University of Mazandaran.
\end{acknowledgements}

\section*{Appendix A}

The $(2+1)$-dimensional action in Einstein-Maxwell-$\Lambda $ gravity is
given 
\begin{equation}
\mathcal{I}(g_{\mu \nu },A_{\mu })=\frac{1}{16\pi }\int_{\partial \mathcal{M}%
}d^{3}x\sqrt{-g}\left[ R-2\Lambda -\mathcal{F}\right],
\end{equation}
where $\mathcal{F}$ is the Maxwell invariant which is equal to $F_{\mu \nu
}F^{\mu \nu }$. It is noteworthy that $F_{\mu \nu }=\partial _{\mu
}A_{\nu}-\partial _{\nu }A_{\mu }$ is the electromagnetic tensor field, and $%
A_{\mu }$ is the gauge potential.

The field equations in this theory of gravity are 
\begin{eqnarray}
G_{\mu \nu }+\Lambda g_{\mu \nu } &=&2\left[ F_{\mu \rho }F_{\nu }^{\rho }+%
\frac{1}{4}g_{\mu \nu }\left( -\mathcal{F}\right) \right],  \label{Max1} \\
&&  \notag \\
\partial _{\mu }\left( \sqrt{-g}F^{\mu \nu }\right) &=&0.  \label{Max2}
\end{eqnarray}

Considering $A_{\mu }=h(r)\delta _{\mu }^{0}$, and the metric (\ref{Metric}%
), and also by considering $\theta =0$, one can show that Eq. (\ref{Max2})
reduces to 
\begin{equation}
\left( 2\alpha r-1\right) h^{\prime }\left( r\right) +r\left( \alpha
r-1\right) h^{\prime \prime }\left( r\right) =0,  \label{dfhMax}
\end{equation}%
solving Eq. (\ref{dfhMax}), one can find $h\left( r\right) $, and the
electric field ($E\left( r\right) $) in the following forms 
\begin{eqnarray}
h\left( r\right) &=&-q\ln \left( \frac{\alpha r-1}{r}\right) ,  \label{hrMax}
\\
&&  \notag \\
E\left( r\right) &=&\frac{-q}{r\left( \alpha r-1\right) },  \label{ErMax}
\end{eqnarray}%
where $q$ is an integration constant related to the electric charge. It is
notable that in the absence of acceleration parameter (i.e., $\alpha =0$),
the electric field turns to standard form ($E\left( r\right) =\frac{q}{r}$),
as we expected.

Now, we are going to find the metric function $f\left( r\right) $.
Considering the obtained $h\left( r\right) $, one can show that the
components of Eq. (\ref{Max1}) reduce to 
\begin{eqnarray}
Eq_{tt} &=&Eq_{rr}=\left( 1-\alpha r\right) rf^{\prime }\left( r\right)
+2\alpha rf\left( r\right) +2\left( \alpha ^{2}\left( q^{2}-m+k\right)
+\Lambda \right) r^{2}  \notag \\
&&  \notag \\
&&-2\left( k-m+2q^{2}\right) \alpha r+2q^{2},  \label{EqttMax1} \\
&&  \notag \\
Eq_{\theta \theta } &=&\left( \alpha r-1\right) ^{2}r^{2}f^{\prime \prime
}\left( r\right) -2\left( \alpha r-1\right) \alpha r^{2}f^{\prime }\left(
r\right) +2\alpha ^{2}r^{2}f\left( r\right) +2\left( \Lambda -\alpha
^{2}q^{2}\right) r^{2}  \notag \\
&&  \notag \\
&&+2q^{2}\left( 2\alpha r-1\right) .  \label{EqttMax2}
\end{eqnarray}

We find an exact solution of Eqs. (\ref{EqttMax1}) and (\ref{EqttMax2}), as 
\begin{equation}
f\left( r\right) =k-m+\left( \left( m-k\right) \alpha ^{2}-\Lambda \right)
r^{2}+2\left( \alpha r-1\right) ^{2}q^{2}\ln \left( \frac{\alpha r-1}{r}%
\right) ,  \label{frMax}
\end{equation}%
where we used $C\left( m,k,\Lambda ,\alpha \right) =\left( m-k\right) -\frac{%
\Lambda }{\alpha ^{2}}$, to find the above solution.

Here we want to check the solution (\ref{frMax}). Without the electrical charge, the solution (Eq. (\ref{frMax})) must recover an uncharged accelerating BTZ black hole. In other words, by replacing $q=0$ in the solution (\ref{frMax}), it reduces to the obtained uncharged accelerating BTZ black hole in the equation (\ref{f(r)Uch}). But we cannot find a real
solution for charged BTZ black hole in the absence of the acceleration parameter, due to the existence of term, $\ln \left( \frac{\alpha r-1}{r}%
\right) $, in the solution (\ref{frMax}). In other words, the term $\ln
\left( \frac{\alpha r-1}{r}\right) $, turns to $\ln \left( \frac{-1}{r}%
\right) $, when $\alpha =0$. So the obtained charged accelerating BTZ black hole from Einstein-Maxwell-$\Lambda $ theory cannot recover the known charged BTZ black hole. This result motivates us to consider the NED field instead of the Maxwell field.

\end{document}